\documentclass{egpubl}
\usepackage{eurovis2021}

\SpecialIssuePaper         %

\usepackage[T1]{fontenc}
\usepackage{dfadobe}

\usepackage{cite}  %
\BibtexOrBiblatex
\electronicVersion
\PrintedOrElectronic
\ifpdf \usepackage[pdftex]{graphicx} \pdfcompresslevel=9
\else \usepackage[dvips]{graphicx} \fi

\usepackage{egweblnk}
\usepackage[disable]{todonotes}

\newcommand{\stodo}[1]{\todo[size=\small, color=yellow]{#1}}
\newcommand{\hntodo}[1]{\todo[size=\small, color=green]{#1}}
\newcommand{\gtodo}[1]{\todo[size=\small, color=pink]{#1}}

\usepackage{xparse}

\usepackage[export]{adjustbox}

\usepackage[percent]{overpic} %
\usepackage{array} %

\usepackage{amsmath}
\usepackage{amssymb}

\usepackage{paralist}   %

\usepackage[all]{nowidow} %

\usepackage{siunitx}
\usepackage[nolist]{acronym}

\DeclareDocumentCommand{\ca}{}{\text{Ca}}
\DeclareDocumentCommand{\can}{m}{\ca=\num{1e#1}}
\DeclareDocumentCommand{\m}{}{\text{M}}
\DeclareDocumentCommand{\mn}{m}{\m=\num{#1}}

\DeclareDocumentCommand{\camn}{mm}{(\can{#1},\mn{#2})}
\DeclareDocumentCommand{\camns}{mm}{(\ca\,\num{1e-#1},\m\,#2)}

\usepackage{microtype}

\usepackage{mathtools}
\DeclarePairedDelimiter\abs{\lvert}{\rvert}%

\usepackage[skip=1pt]{subcaption}
\definecolor{ca5m10}{rgb}{0.921569,0.631373,1}
\definecolor{ca4m10}{rgb}{0.921569,0.631373,0.733333}
\definecolor{ca3m10}{rgb}{0.921569,0.631373,0.466667}
\definecolor{ca2m10}{rgb}{0.921569,0.631373,0.0117647}
\definecolor{ca5m1}{rgb}{0.560784,0.596078,1}
\definecolor{ca4m1}{rgb}{0.560784,0.464052,0.733333}
\definecolor{ca3m1}{rgb}{0.560784,0.415686,0.466667}
\definecolor{ca2m1}{rgb}{0.560784,0.415686,0.105882}
\definecolor{ca5m02}{rgb}{0,0.596078,1}
\definecolor{ca4m02}{rgb}{0.0666667,0.464052,0.733333}
\definecolor{ca3m02}{rgb}{0.133333,0.332026,0.466667}
\definecolor{ca2m02}{rgb}{1,0,0}

\title[Visual Analysis of Flow Displacement in Porous Media]%
      {Visual Analysis of Two-Phase Flow Displacement Processes\\ in Porous Media}

\author[S. Frey, S. Scheller, N. Karadimitriou, D. Lee,  G. Reina, H. Steeb, T. Ertl]
       {\parbox{\textwidth}{\centering S. Frey$^{1}$, S. Scheller$^{2}$, N. Karadimitriou$^{2}$, D. Lee$^{2}$, G. Reina$^{2}$, H. Steeb$^{2}$, T. Ertl$^{2}$
                      }
                      \\
                        {\parbox{\textwidth}{\centering $^1$University of Groningen, Netherlands\\
                            $^2$University of Stuttgart, Germany                            
                          } 
                        }
                    }

      \DeclareDocumentCommand{\nframe}{mm}{figs/20200429_networks/Ca=10#1,M=#2/transport-network/before_breakthrough}
      \def\nImgWidth{0.232\linewidth}
      
\begin{document}
      \teaser{
        \begin{minipage}[b][][b]{0.72\linewidth}

          \subcaptionbox*{\label{fig:grid_before_breakthrough:ca5m02}\color{ca5m02} $\ca=\num{1e-5}, \m=0.2$}%
                         {\includegraphics[width=\nImgWidth,cfbox=ca5m02 2pt 0pt]{\nframe{-5}{0.2}}}
                         \hfill
                         \subcaptionbox*{\label{fig:grid_before_breakthrough:ca4m02}\color{ca4m02} $\ca=\num{1e-4}, \m=0.2$}
                                        {\includegraphics[width=\nImgWidth,cfbox=ca4m02 2pt 0pt]{\nframe{-4}{0.2}}}
                                        \hfill
                                        \subcaptionbox*{\label{fig:grid_before_breakthrough:ca3m02}\color{ca3m02} $\ca=\num{1e-3}, \m=0.2$}
                                                       {\includegraphics[width=\nImgWidth,cfbox=ca3m02 2pt 0pt]{\nframe{-3}{0.2}}}
                                                       \hfill
                                                       \subcaptionbox*{}
                                                                      {\phantom{\includegraphics[width=\nImgWidth,cfbox=ca3m02 2pt 0pt]{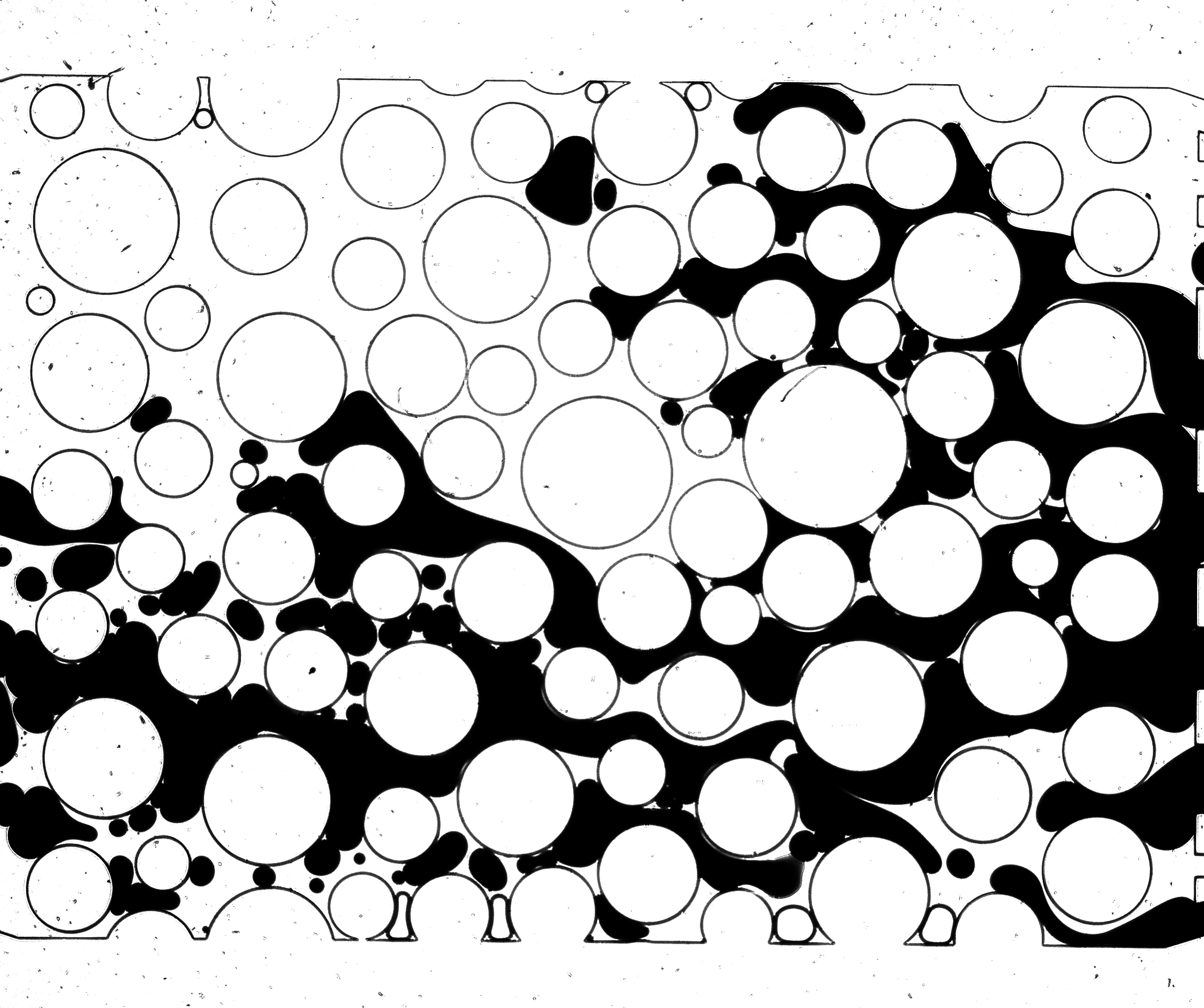}}}

                                                               \subcaptionbox*{\label{fig:grid_before_breakthrough:ca5m1}\color{ca5m1} $\ca=\num{1e-5}, \m=1$}%
                                                                             {\includegraphics[width=\nImgWidth,cfbox=ca5m1 2pt 0pt]{\nframe{-5}{1}}}
                                                                             \hfill
                                                                             \subcaptionbox*{\label{fig:grid_before_breakthrough:ca4m1}\color{ca4m1} $\ca=\num{1e-4}, \m=1$}
                                                                                           {\includegraphics[width=\nImgWidth,cfbox=ca4m1 2pt 0pt]{\nframe{-4}{1}}}
                                                                                           \hfill
                                                                                           \subcaptionbox*{\label{fig:grid_before_breakthrough:ca3m1}\color{ca3m1} $\ca=\num{1e-3}, \m=1$}
                                                                                                         {\includegraphics[width=\nImgWidth,cfbox=ca3m1 2pt 0pt]{\nframe{-3}{1}}}
                                                                                                         \hfill
                                                                                                         \subcaptionbox*{\label{fig:grid_before_breakthrough:ca2m1}\color{ca2m1} $\ca=\num{1e-2}, \m=1$}
                                                                                                                       {\includegraphics[width=\nImgWidth,cfbox=ca2m1 2pt 0pt]{\nframe{-2}{1}}}

                                                                                                                       \subcaptionbox*{\label{fig:grid_before_breakthrough:ca5m10}\color{ca5m10} $\ca=\num{1e-5}, \m=10$}%
                                                                                                                                      {\includegraphics[width=\nImgWidth,cfbox=ca5m10 2pt 0pt]{\nframe{-5}{10}}}
                                                                                                                                      \hfill
                                                                                                                                      \subcaptionbox*{\label{fig:grid_before_breakthrough:ca4m10}\color{ca4m10} $\ca=\num{1e-4}, \m=10$}
                                                                                                                                                     {\includegraphics[width=\nImgWidth,cfbox=ca4m10 2pt 0pt]{\nframe{-4}{10}}}
                                                                                                                                                     \hfill
                                                                                                                                                     \subcaptionbox*{\label{fig:grid_before_breakthrough:ca3m10}\color{ca3m10} $\ca=\num{1e-3}, \m=10$}
                                                                                                                                                                    {\includegraphics[width=\nImgWidth,cfbox=ca3m10 2pt 0pt]{\nframe{-3}{10}}}
                                                                                                                                                                    \hfill                                                                                                                                                                    
                                                                                                                                                                    \subcaptionbox*{\label{fig:grid_before_breakthrough:ca2m10}\color{ca2m10} $\ca=\num{1e-2}, \m=10$}
                                                                                                                                                                                   {\includegraphics[width=\nImgWidth,cfbox=ca2m10 2pt 0pt]{\nframe{-2}{10}}}

       \end{minipage}
       \hfill
       \subcaptionbox*{color maps for pores and throats}{%
  \begin{minipage}[b][][b]{0.0735\textwidth}
    \includegraphics[width=\textwidth, height=6.5484\textwidth]{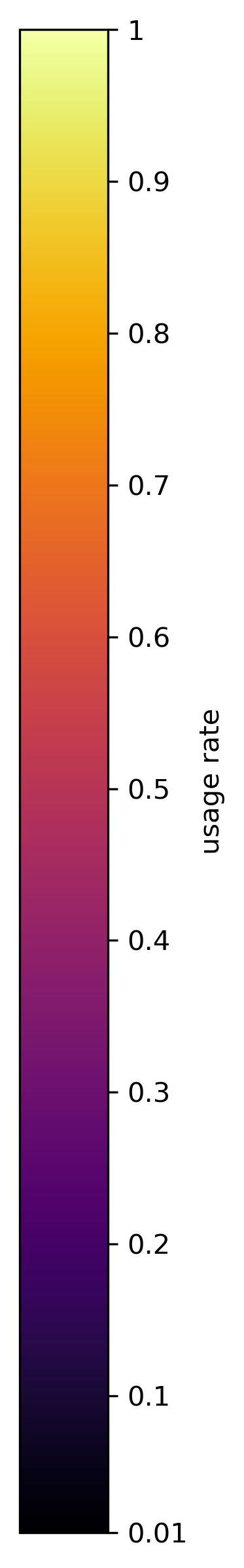}
  \end{minipage}
  \begin{minipage}[b][][b]{0.15\textwidth}
    \includegraphics[width=\textwidth, height=3.22\textwidth]{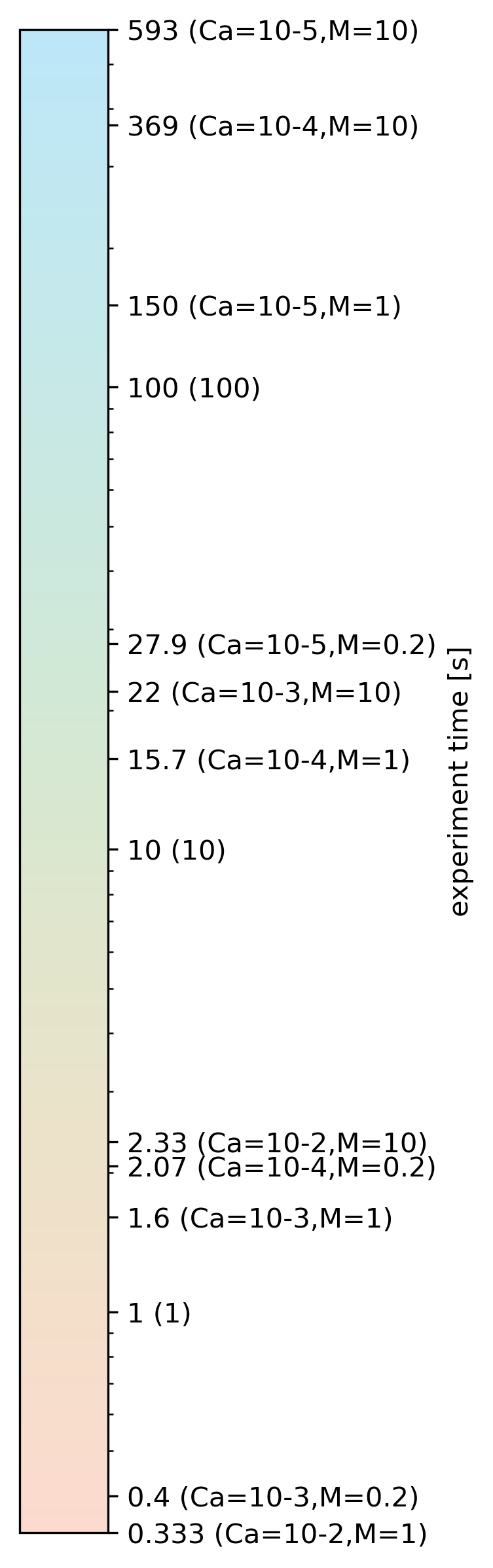}
  \end{minipage}}
  \caption{%
    Transport paths in the porous medium for varying capillary numbers $\ca{}$ and viscosity ratios $\m{}$ in drainage experiments~(i.e.,~displacement of the fluid---the wetting phase---initially filling the solid with another fluid---the non-wetting phase---entering from the right).
    Nodes represent pore bodies (diameter$\hat{=}$size), links stand for pore throats (edge thickness$\hat{=}$width).
    Experiments capture different time scales that are made comparable in this visualization by considering the time span between (i) the non-wetting fluid entering and (ii) forming an uninterrupted connection from left to right~(aka breakthrough).
    Usage rate $\mathcal{U}$---a normalized measure of how long a pore or throat has been occupied by the non-wetting phase--- is mapped to color directly (for $\mathcal{U}\approx 0$ the duration of the experiment is shown instead).
    This allows to draw conclusions regarding dominant driving forces of the flow and detect unexpected events.
    Frames around each experiment depict its representative color used throughout this paper.    
  }
  \label{fig:grid_before_breakthrough}  
}

\maketitle
\begin{abstract}
  We present the visual analysis of our novel parameter study of porous media experiments, focusing on gaining a better understanding of drainage processes on the micro-scale.
  We analyze the temporal evolution of extracted characteristic values, and discuss how to directly compare experiments that exhibit processes at different temporal scales due to varying boundary and physical conditions.
  To enable spatio-temporal analysis, we introduce a new abstract visual representation showing which paths through the porous media were occupied to what extent, e.g., allowing for classification into viscous and capillary regimes.
  This joint work of porous media experts and visualization researchers yields new insights regarding immiscible two-phase flow on the micro-scale toward the overarching goal of characterizing flow based on boundary conditions and physical fluid properties.


\end{abstract}

\begin{acronym}
  \acro{PDMS}{Poly-Di-Methyl-Siloxane}
  \acro{PTFE}{Poly-Tetra-Fluoro-Ethylene}
\end{acronym}

\section{Introduction}
\label{sec:intro}

\hntodo{TODO for Holger and Nikos}
\stodo{TODO for Steffen}
\gtodo{TODO for Guido}

A thorough understanding of the underlying physical processes occurring during two-phase flow is of high importance, since it will enhance our abilities to effectively build models for prediction and simulation.
Here, two-phase flow term refers to the interactive flow of two fluids in a confined space surrounded by solid walls.
The two fluids interact via their common interfaces, and they have no or very limited mixing potential with each other.
Based on this assumption, at least one of the fluids has to be a liquid so that immiscible flow can take place. The fluid which has the stronger affinity to the solid phase (increased spreading) is usually referred to as the wetting phase, while the fluid with the least affinity is referred to as the non-wetting phase.
The processes involved, like drainage (where the non-wetting phase forcibly displaces the wetting phase) and imbibition (the inverse process which can also take place without external force), have been the topic of extensive work in porous media research due to the degree of degeneracy that is expressed via hysteresis in the capillary pressure-saturation relation~\cite{jerauld1990effect,hilfer2006capillary}, and how this hysteresis can be lifted~\cite{hassanizadeh1993thermodynamic}.
Such fundamental physical processes play a crucial role in the handling and optimization of industrial and environmental applications with a dominant socio\-economic impact, including enhanced oil recovery~\cite{alzahid2017}, soil remediation~\cite{khan2004}, printing~\cite{kettle2010}, etc.
Given the fundamental significance of two-phase flow, numerical~\cite{joekar2010non,porter2009lattice,bryant1992prediction} and experimental~\cite{wildenschild2013x,doi:10.1029/97WR02115} schemes have been developed to describe the underlying physical processes in the most efficient way given the complexity of the process and the physical constrains to acquire pore-scale experimental evidence.
However, a better understanding of the complex dynamic interplay between viscous and capillary forces during a drainage event, and the impact of the pore space geometry, boundary conditions, and physical properties of the involved fluids is crucial for further advancements in the field.

The close collaboration between porous media researchers and visualization experts presented in this work yielded new insights toward the overarching goal of flow characterization based on the correlation of boundary conditions and physical properties of the fluids in a porous medium.
We argue that this lays a crucial foundation for further advancements in the domain.
For this, we conducted and analyzed eleven two-phase displacement experiments, employing an artificial, Poly-Di-Methyl-Siloxane (PDMS) micro-model under varied viscosity ratios~$\m{}$~ between the invading and the defending fluid under a drainage scenario, and capillary numbers~$\ca{}$~(which provides a measure of the competition of the viscous vs the capillary forces for the corresponding boundary conditions).
\hntodo{DO WE ACTUALLY DO THIS IN THIS PAPER?  , in an attempt to reproduce the concept from Lenormand et al. \cite{lenormand1988numerical}, which will be elaborated soon after.}
In this work, we aim at linking the flow behaviour to $\ca$ and $\m$, with the ultimate objective being the inclusion of the local porous medium geometry as a determining measure.
This is expected to enable the investigation of how and why flow characteristics can change with respect to the heterogeneity and the local geometrical properties of the pore space.
The ability to visualize experiments in a compact way and to quantify values of interest allows us to identify similarities and differences between the processes, to categorize the experimental results and to further research the relation between boundary conditions, physical fluid properties and the resulting flow.
This requires the extraction and expressive presentation of relevant quantities, which also need to be made directly comparable as---depending on the viscosity ratio~$\m$ and capillary number~$\ca$---the speed of flow progression is in different orders of magnitude~(ranging between \SI{330}{\milli\second} and \SI{593}{\second}, cf.~\autoref{fig:grid_before_breakthrough}).
In particular, we visually analyze the occupancy of the pore space by each phase, the breakup of the connected phases due to the competition between viscous and interfacial effects, and coalescence of disconnected blobs, either to form bigger blobs or to reconnect with the continuous phase, and the production of interfacial area and the mechanisms behind that.
\hntodo{can we rightfully claim that we do all that in this paper?}
As the first work in the domain, we also study how these parameters correlate to the pore space and the local geometrical properties.
In the following, we first review related work in visualization as well as porous media research~(\autoref{sec:related}), before describing our experiments and the extraction of phases~(\autoref{sec:experiments}).
This provides the basis for determining visualization data for comparable analysis, including the extraction of quantities of interest (e.g., interfaces, saturation, inlet connection), and accounting for different time scales across experiments via temporal normalization~(\autoref{sec:comparability}).
This then allows to conduct a quantitative comparative analysis of time-dependent processes~(\autoref{sec:charactersitic_values}).
For spatial analysis in the pore space, we employ an aggregate network representation of the porous structure that allows to investigate which and to what degree paths through the porous medium were taken by a fluid~(\autoref{sec:networks}).
Finally, we summarize the impact of this work on porous media research and discuss limitations as well as directions for future work~(\autoref{sec:discussion}).

\section{Related Work in Visualization and Porous Media Research}
\label{sec:related}

\textbf{Visualization of porous media and liquid phases.}
Earlier work focused mostly on the rendering of porous media: Grottel et al.~\cite{grottel_particle-based_2010} show different approaches employing geometry shaders or ray casting, while Naumov et al.~\cite{naumov_rendering_2013} investigate rendering flow in a porous medium in a VR environment and tackling occlusion issues.
Zhang et al.~\cite{zhang2018visualization} extracted CO\textsubscript{2} bubbles and their surrounding structure in a liquid-filled sandstone sample from X-ray computed tomography data.
They automatically classify and correlate bubbles with similar interface morphology and geometric features and support the search for porous structures that favor the sequestration of CO\textsubscript{2} bubbles.
De Winter et al.~\cite{winter:20} investigate the behavior of single-phase flow in microfluidic structures from experimental recordings via confocal laser scanning.
Among others, they visualize the boundaries between different flow regions and depict transport mechanisms via streamlines and animated renderings.
The Scientific Visualization Contest 2016 featured a particle simulation ensemble of salt dissolving in water, developing viscous fingers~\cite{contest:16}.
This data set was the focus of several approaches, also beyond the contest scope.
Gralka et al.~\cite{gralka:18} describe a system for the visual and structural investigation of this data via multiple views, drilling down from diagrams of ensemble metrics to investigation of the 3D data, abstracted finger topology and vortex core lines forming in the data.
Favelier et al.~\cite{favelier2016visualizing} extract and track the emerging fingers based on topological data analysis and visualize their properties in comparative line plots.
This work was extended to include tracking graphs~\cite{widanagamaachchi_interactive_2012} to investigate a single simulation and a query interface coupled to thumbnails of a Paraview Cinema database to investigate the ensemble as a whole~\cite{lukasczyk2017viscous}.
An alternative approach complements the 3D and structure views by glyphs summarizing high-level parameters of single members of the ensemble~\cite{luciani2018details}.
Soler et al.\cite{soler2019ranking} rank an ensemble of viscous fingering simulations according to their agreement with a ground-truth data set.
To this end they propose a novel metric that combines geometric and topological features based on persistence diagrams.
Xu et al.~\cite{xu2019geometry} use Reeb graphs to extract viscous and gravitational finger structures and the respective skeletons, which are  abstracted into glyphs and tracked over time.
They also demonstrate a tracking graph augmented with the glyphs as well as an analysis system which allows for interactively browsing through the spatial domain as well as the extracted fingers.

\textbf{Ensemble visualization.}
The analysis of ensemble data generally is a challenging visualization task~\cite{g:obermaier:2014}.
Potter et al.~\cite{g:potter:2009} as well as Sanyal et al.~\cite{g:sanyal:2010} proposed early approaches to study climate ensembles, while Waser et al.~\cite{g:waser:2010} described a system for the interactive steering of simulation ensembles.
Kehrer et al.~\cite{g:kehrer:2013}, Sedlmair~et~al.~\cite{g:sedlmair:2014}, and Wang~et~al.~\cite{g:wang:2019a} provided detailed surveys of techniques in the area.
Bruckner and~M\"{o}ller~\cite{g:bruckner:2010} employ squared differences to explore the visual effects simulation space, Hummel~et~al.~\cite{g:hummel:2013} compute region similarity via joint variance, and Kumpf~et~al.~\cite{g:kumpf:2019} track statistically-coherent regions using optical flow.
Hao~et~al.~\cite{g:hao:2016a} calculate shape similarities for particle data using an octree structure, while He~et~al.~\cite{g:he:2020} employ surface density estimates for distances between surfaces.

\textbf{Positioning in Porous Media Research.}
Various macroscopical continuum models for immiscible two-phase flow in porous media have been proposed and widely applied.
Van Genuchten~\cite{van1980closed} introduced a "Darcy-scale" model which proposed that the capillary pressure is a simplified non-linear relation of the wetting phase saturation.
Based on thermodynamical principles and approaches, it has been shown that capillary pressure does not only depend on the fluids' saturation, but crucially also on its spatial distribution in the porous domain\cite{schluter2016pore}.
Numerous attempts have been made to embed the pore-scale geometry and the spatial fluids' configuration in the pore space in continuum models.
One of these is based on a multi-scale approach that includes the specific interfacial area between the two phases as a separate state variable in addition to phase saturation\cite{hassanizadeh1993thermodynamic}.
Such theories including additional state variables are commonly denoted as "extended theories".
They can also include process variables dependent on higher-order morphological features\cite{kurzeja2014variational} which can be expressed with the first three Minkowski functionals \textit{M0 –M2}\cite{webster2001statistical}.
One common underlying assumption is that none of the two phases gets disconnected during the process.
However, this only holds for ideal conditions which are hardly ever observed in practice, even in very well-conditioned experiments for drainage and imbibition~(like the one discussed in this paper).

\textbf{Micro-model experiments.}
Two-phase flow has been studied in experiments by means of natural or artificial porous media.
As natural porous media mostly rock samples are used~\cite{bartels2016micro,Berg3755,doi:10.1002/grl.50771}, while for the latter there are many approaches to manufacture porous media with well-defined properties~\cite{darcy1857recherches,grosser1988onset,shaw1898investigation,doi:10.1029/97WR02115}.
In this work, the artificial porous media used are PDMS micromodels.
As first defined by Karadimitriou and Hassanizadeh\cite{doi:10.2136/vzj2011.0072}, the term "micromododels"" is used to describe an artificial, transparent porous medium with similar average properties, like pore size distribution, porosity and permeability, to a natural porous medium, with the pores measuring on the micro-scale and the overall dimensions on the centimeter scale.
Micro-models are commonly produced either by etching processes\cite{mattax1961ever}, or lithography\cite{thompson1983introduction}.
Soft-lithography specifically is widely employed due to its comparably low cost and the high precision with which copies of the same flow network can be made\cite{XiaWhitesides1998}.
As this type of micro-models is transparent, images taken during flow and displacement can be acquired in high temporal and spatial resolution with optical means.

In the literature there is a significant volume of works which make use of micro-models and optical illumination in order to visualize the effects taking place, at a spatial resolution of a few microns per sensor pixel.
With such a micromodel, Lenormand et al.\cite{lenormand1988numerical} investigated the effect of the boundary conditions and the physical properties of the involved fluids, both numerically and experimentally.
They performed drainage experiments for a combination of boundary conditions and physical fluid properties and created a well-known map of the parameter space and the resulting displacement regimes.
They categorized the flow regimes into \emph{capillary fingering}, \emph{viscous fingering} as well as \emph{stable front} and were able to match these regimes both numerically and experimentally.
Probably due to the scarcity of computational resources and image analysis tools they limited themselves to qualitative observations of the experimental images, without being able to draw any quantitative conclusions.
Cheng et al.\cite{doi:10.1029/2003GL019282} performed drainage and imbibition experiments, in an attempt to investigate the role of specific interfacial area as a separate state variable in two-phase flow studies, under quasi-static flow conditions.
This is the first experimental proof---via recording features on the micro-scale---that specific interfacial area can potentially be used as a state variable in two-phase flow studies.
Karadimitriou et al.\cite{doi:10.1002/2014WR015388} discuss the possibility of the inclusion of specific interfacial area as a state variable for two-phase flow, but under dynamic conditions.
They used a micro-model based on \ac{PDMS} and akin to Cheng et al.\cite{doi:10.1029/2003GL019282} they performed drainage and imbibition experiments, including scanning curves, for three different fluxes, ranging from a capillary to a highly viscous flow regime.
They extracted information on phase saturation, local capillary pressure, local contact angle, location of the apex of the interfaces and constructed again the corresponding surface in the \emph{capillary pressure}--\emph{saturation}--\emph{specific interfacial area} space.
They concluded that specific interfacial area can be included as a state variable in two-phase flow studies under quasi-static but not under dynamic conditions.
While they attributed the observed mismatch mostly to the disconnected phases induced during the process, they could not quantify the disconnected phases in terms of saturation and specific interfacial area (which we do in this work).

\begin{figure}[t]
  \adjincludegraphics[width = \linewidth, trim={{0.0\width} {0.00\height} {0.\width} {0.2\height}}, clip]{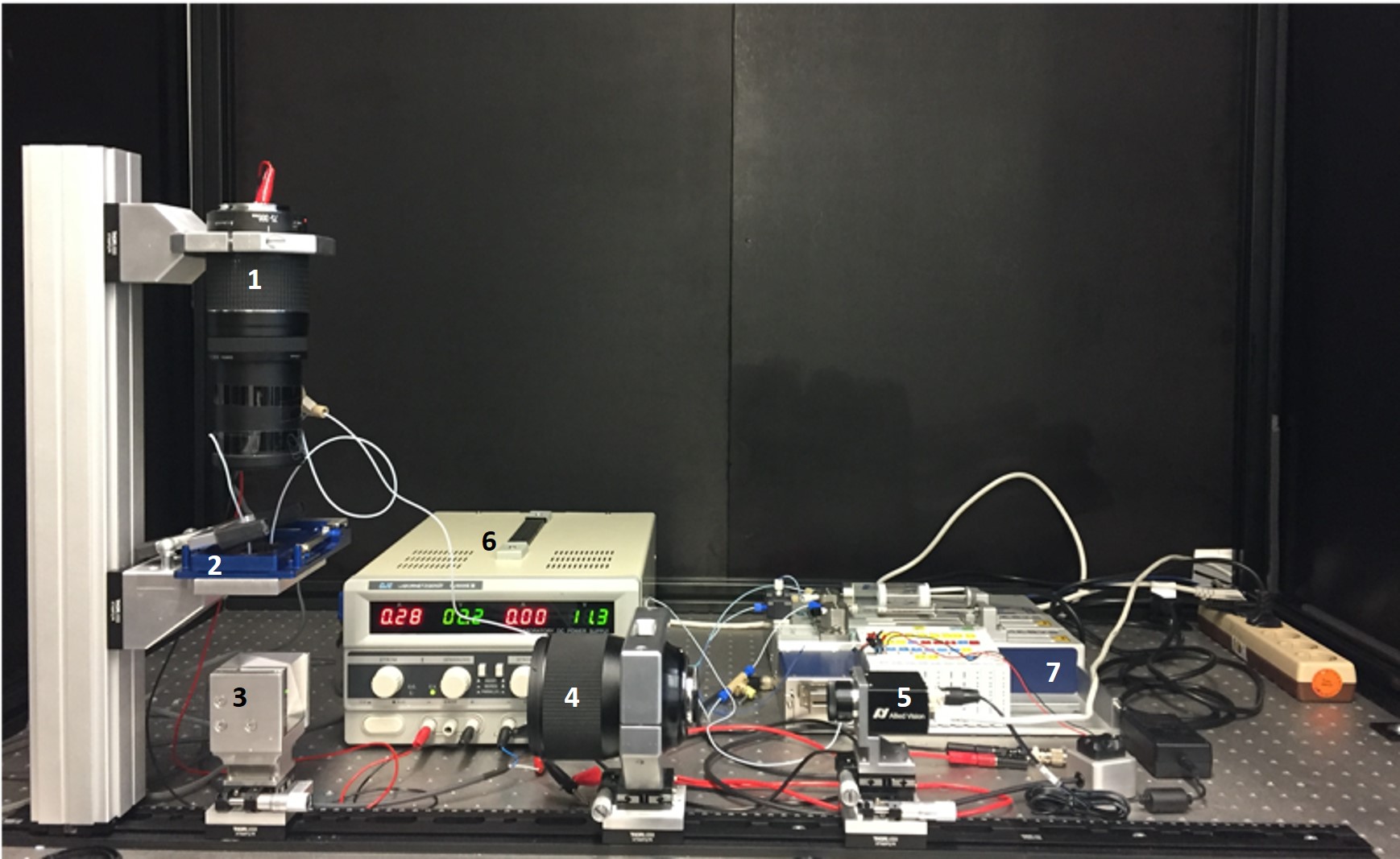}
  \caption{Optical setup capturing the experiments: (1)~LED source mounted to an objective lens, (2)~the stage for the micro-model,
    (3)~a prism, (4)~a telephoto lens, (5)~an industrial camera (connected to a PC via Ethernet), (6)~a stable power supply, and (7)~the syringe pumps.
    The setup was placed on a vibration-free optical table.
  }
  \label{fig:opticalsetup}
\end{figure}

\DeclareDocumentCommand{\bframe}{mm}{figs/breakthroughFrames/Ca=10#1,M=#2}
\def\expImgWidth{0.218\linewidth}
\begin{figure}
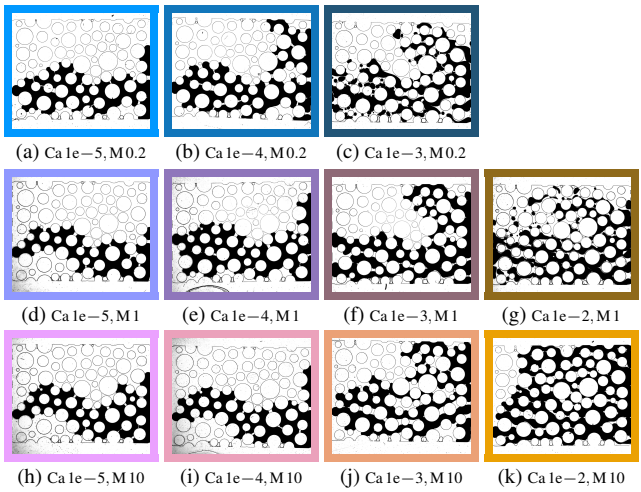

  \subcaptionbox{\label{fig:expimgs:ca5m02}\tiny $\ca\,\num{1e-5}, \m\,0.2$}%
                {\includegraphics[width=\expImgWidth,cfbox=ca5m02 3pt 0pt]{\bframe{-5}{0.2}}}
                \hfill
                \subcaptionbox{\label{fig:expimgs:ca4m02}\tiny $\ca\,\num{1e-4}, \m\,0.2$}
                              {\includegraphics[width=\expImgWidth,cfbox=ca4m02 3pt 0pt]{\bframe{-4}{0.2}}}
                              \hfill
                              \subcaptionbox{\label{fig:expimgs:ca3m02}\tiny $\ca\,\num{1e-3}, \m\,0.2$}
                                            {\includegraphics[width=\expImgWidth,cfbox=ca3m02 3pt 0pt]{\bframe{-3}{0.2}}}
                                            \hfill
                              \subcaptionbox*{}
                                             {\phantom{\includegraphics[width=\expImgWidth,cfbox=ca3m02 3pt 0pt]{figs/plots/cluster/Ca=10-3_M=0d2_1345.png}}}

                                             \subcaptionbox{\label{fig:expimgs:ca5m1}\tiny $\ca\,\num{1e-5}, \m\,1$}%
                                                           {\includegraphics[width=\expImgWidth,cfbox=ca5m1 3pt 0pt]{\bframe{-5}{1}}}
                                                           \hfill
                                                           \subcaptionbox{\label{fig:expimgs:ca4m1}\tiny $\ca\,\num{1e-4}, \m\,1$}
                                                                         {\includegraphics[width=\expImgWidth,cfbox=ca4m1 3pt 0pt]{\bframe{-4}{1}}}
                                                                         \hfill
                                                                         \subcaptionbox{\label{fig:expimgs:ca3m1}\tiny $\ca\,\num{1e-3}, \m\,1$}
                                                                                       {\includegraphics[width=\expImgWidth,cfbox=ca3m1 3pt 0pt]{\bframe{-3}{1}}}
                                                                                       \hfill
                                                                                       \subcaptionbox{\label{fig:expimgs:ca2m1}\tiny $\ca\,\num{1e-2}, \m\,1$}
                                                                                                     {\includegraphics[width=\expImgWidth,cfbox=ca2m1 3pt 0pt]{\bframe{-2}{1}}}

                                                                                                     \subcaptionbox{\label{fig:expimgs:ca5m10}\tiny $\ca\,\num{1e-5}, \m\,10$}%
                                                                                                                   {\includegraphics[width=\expImgWidth,cfbox=ca5m10 3pt 0pt]{\bframe{-5}{10}}}
                                                                                                                   \hfill
                                                                                                                   \subcaptionbox{\label{fig:expimgs:ca4m10}\tiny $\ca\,\num{1e-4}, \m\,10$}
                                                                                                                                 {\includegraphics[width=\expImgWidth,cfbox=ca4m10 3pt 0pt]{\bframe{-4}{10}}}
                                                                                                                                 \hfill
                                                                                                                                 \subcaptionbox{\label{fig:expimgs:ca3m10}\tiny $\ca\,\num{1e-3}, \m\,10$}
                                                                                                                                               {\includegraphics[width=\expImgWidth,cfbox=ca3m10 3pt 0pt]{\bframe{-3}{10}}}
                                                                                                                                               \hfill
                                                                                                                                               \subcaptionbox{\label{fig:expimgs:ca2m10}\tiny $\ca\,\num{1e-2}, \m\,10$}
                                                                                                                                                             {\includegraphics[width=\expImgWidth,cfbox=ca2m10 3pt 0pt]{\bframe{-2}{10}}}

  \caption{
    Experiment recordings at breakthrough~(an uninterrupted connection exists between left and right boundary, \autoref{sec:tempnorm}).
    }
  \label{fig:expimgs}
  \label{fig:expbreak}
\end{figure}

\begin{figure}[t]
  \centering
  \subcaptionbox{\label{fig:mask_processing_original}original capture}{\includegraphics[width=0.3\columnwidth]{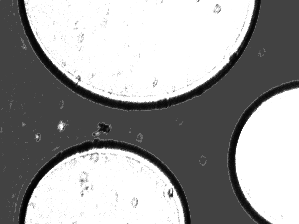}}
  \hfill
  \subcaptionbox{\label{fig:mask_processing_tweaked}manual cleaning}{\includegraphics[width=0.3\columnwidth]{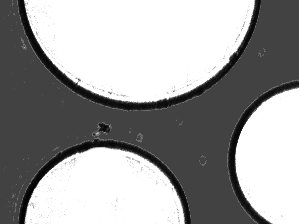}}
  \hfill
  \subcaptionbox{\label{fig:mask_processing_final}post-processing}{\includegraphics[width=0.3\columnwidth]{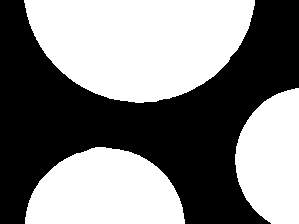}}
  \caption{Mask zoom-in of experiment $(\ca=\num{1e-5}, \m = 1)$.
  }
  \label{fig:mask_processing}
\end{figure}

\def\segImgWidth{0.16\linewidth}
\begin{figure*}[t]
  \subcaptionbox{\label{fig:segmentation_mask} mask}{\includegraphics[width=\segImgWidth]{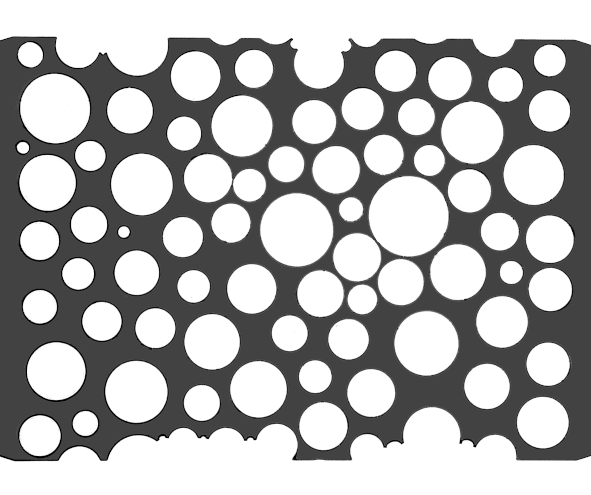}}
  \hfill
  \subcaptionbox{\label{fig:segmentation_input} captured frame}{\includegraphics[width=\segImgWidth]{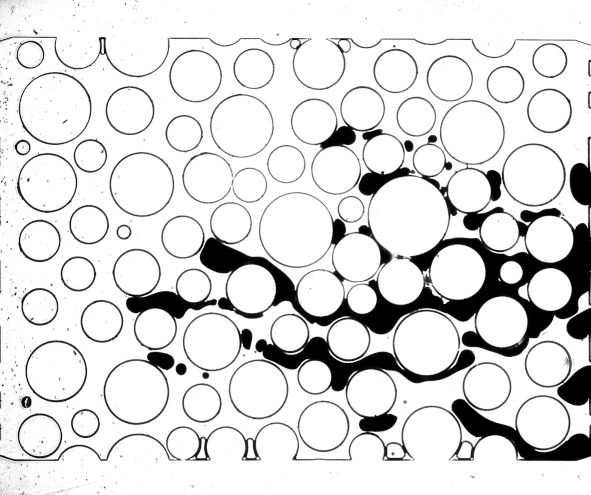}}
  \hfill
\subcaptionbox{\label{fig:segmentation_segmented}phases}{\includegraphics[width=\segImgWidth]{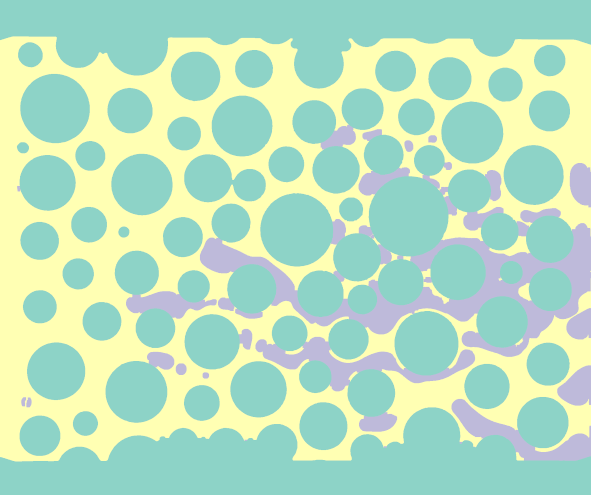}}
\hfill
\subcaptionbox{\label{fig:segmentation_interface}interfaces}{\includegraphics[width=\segImgWidth]{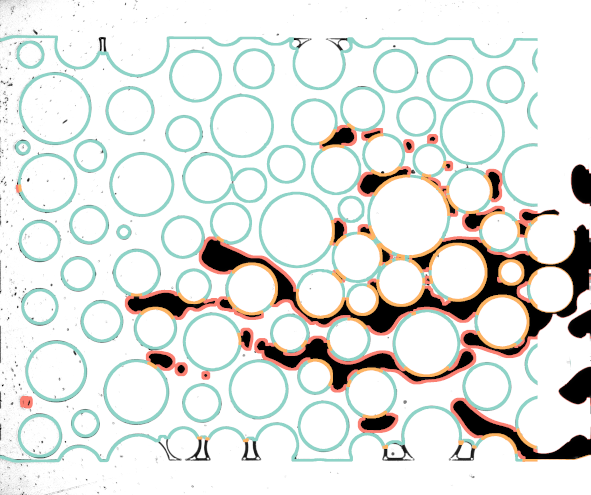}}
\hfill
\subcaptionbox{\label{fig:segmentation_segmented_detailed}connected phases}{\includegraphics[width=\segImgWidth]{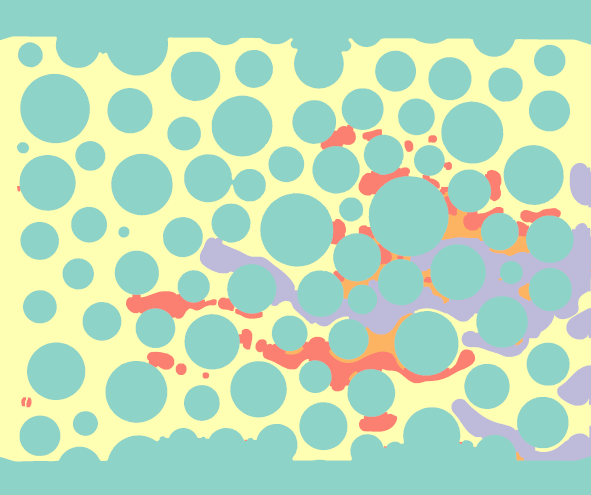}}
\hfill
\subcaptionbox{\label{fig:segmentation_interface_detailed}connected interfaces}{\includegraphics[width=\segImgWidth]{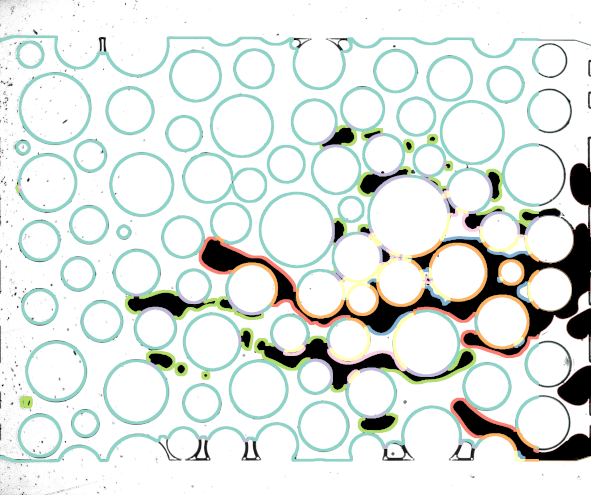}}
  \caption{%
    Phase segmentation and interface extraction example for $(\ca{}=\num{1e-2}$, $\m{}=1)$.
    (\subref{fig:segmentation_mask})~Using the mask, (\subref{fig:segmentation_input})~captured images are (\subref{fig:segmentation_segmented})~segmented into solid, wetting and non-wetting phase~(cf.~\autoref{sec:imageanalysis}).
    (\subref{fig:segmentation_interface})~We further compute interfaces between all phase combinations.
    (\subref{fig:segmentation_segmented_detailed},\subref{fig:segmentation_interface_detailed})~The non-wetting phase is finally partitioned into parts with and without uninterrupted connection to the inlet~(cf.~\autoref{sec:derived_values}).
  }
  \label{fig:segmentation}
\end{figure*}

\section{Experiments and Image Analysis}
\label{sec:experiments}

We conducted experiments to study the distribution of the two fluid phases depending on the flow boundary conditions and the change in the physical properties of the fluids in combination with the pore geometry~(\autoref{sec:experimentsetup}).
Initially, the pore space is occupied by a wetting phase, which is then displaced by a non-wetting phase during the course of the experiment (this scenario is also known as primary drainage).
The three involved phases---solid, wetting fluid, and non-wetting fluid---are extracted from the resulting time-dependent ensemble for further analysis~(\autoref{sec:imageanalysis}).
While our experimental setup allows for the comparably simple distinction between phases in general, high-precision segmentation is challenging.
In particular, trapped air and dust particles and model imperfections can lead to sizeable deviations that require manual correction to complement automatic processing.

\subsection{Experiments}
\label{sec:experimentsetup}

All experiments use an artificial porous medium made of PDMS via soft lithography \cite{XiaWhitesides1998,Karadimitriouetal2013}, replicating the C-type network used in the work of Sivanesapillai and Steeb\cite{sivanesapillai2018fluid}.
The flow network has a mean pore size of \SI{410}{\micro\metre}, and a porosity of 0.44~(as a comparative measure between the available space for flow and the bulk volume of the porous medium).
A primary drainage scenario is considered in which the flow network that is initially filled with the wetting phase (Fluorinert FC-43) is invaded by the non-wetting phase (water dyed with ink), as mentioned earlier.
To achieve higher viscosities for the invading phase, thus increasing the viscosity ratio, water is mixed with glycerol at given concentrations so as to achieve viscosities equal to and ten times higher than this of Fluorinert.
A \SI{1}{\milli\liter} glass syringe in combination with a mid-pressure CETONI neMESYS 100N~\cite{nemesys} generates the flow in the network via an 1/16~inch ID \ac{PTFE} tube.
Right before the inlet of the flow network, an MPS2~\cite{mps2} flow-through pressure sensor (maximum pressure 1 bar) measures the inlet pressure continuously.
Another identical sensor was also used right after the outlet of the flow network.
With this, pressure drop across the whole network can be measured at a frequency of \SI{10}{\hertz}.

The transparent nature of \ac{PDMS} allows capturing  the processes taking place in the pore space in real time, by using transmitted light microscopy~(\autoref{fig:opticalsetup}).
For this purpose, a custom made microscope was developed, which is able to visualize samples with a resolution of 0.5 to 20 microns per camera pixel.
In this case, flow can be captured at a resolution of 8 microns/pixel, at a frame rate varying from 1 to 15 fps, depending on the speed of the investigated process indicated by the capillary number.
A total of eleven experiments was conducted; the flow boundary conditions included capillary numbers $\ca\in\{\num{1e-5}, \num{1e-4}, \num{1e-3}, \num{1e-2}\}$, and the viscosity ratios were $\m\in\{0.2, 1, 10\}$.
For the combination of $\ca=\num{1e-2}$ and $\m=0.2$, capturing the processes reliably was impossible due to (i)~the very high flow velocity that (ii)~is also sufficient to deform the pore space.
Accordingly, this yields irreproducible results and naturally prohibits meaningful comparison to the other experiments.

A gray-scale, 8-bit image sequence is recorded for every experiment~(e.g.,~\autoref{fig:expimgs})
The wetting phase, \ac{PDMS}, and some trapped air appear as light-gray to white pixels, while the non-wetting fluid, interfaces and dust particles appear as dark-gray to black pixels.
The micro-model is flushed between experiments to remove residual liquids. %
This leads to translational and rotational displacements in the recorded image sections as well as slight perspective changes.
To compensate for this, we capture one mask image for each experiment after saturating the micro-model with the wetting phase (here, the wetting phase appears in dark gray, cf.~\autoref{fig:mask_processing_original}), and employ the \emph{Enhanced Correlation Coefficient (ECC)} for correction~\cite{evangelidis2008parametric}.
For this, we compute a homography between the mask of the experiment with $(\ca = \SI{1e-2}, \m = 1)$ and the masks of all other experiments.
The mask is further used as reference for distinguishing the three phases in the input images below, representing the pore space.
\subsection{Phase Segmentation}
\label{sec:imageanalysis}
\label{sec:values}

Phase segmentation is conceptually straight-forward: masks allow to distinguish between solid and fluid phases, and the experimental setup already relatively clearly distinguishes fluids as light and dark gray.
However, automatic phase segmentation would erroneously recognize non-existent structures and the detected interfaces are faulty.
Note that accurate masks are particularly important: besides segmentation, masks also crucially provide the basis to create the structure of transport networks~(cf.~\autoref{sec:network_generation}\,\textbf{(1)}).
In the recorded masks, the walls of the \ac{PDMS} structure appear as $\approx 10$ pixel wide contours with a dark-gray to black tone that becomes lighter towards the edge~(cf.~closeup in~\autoref{fig:mask_processing_original}).
The contours exhibit noise in brightness and thickness, which is caused by the fact that the cross-sectional shape of the solid edges in ``depth'' in 3D are not precisely rectangular, and the illumination is not perfectly homogeneous.
Additional artifacts are caused by dust particles from the air, which deposit on the model or the optical setup.
They appear as stains with different gray values in the solid or fluid phase and their contours might interfere with those of the \ac{PDMS} surface.
Accordingly, flow paths may not be recognized in areas where there is only a gap of a few microns in the porous media between the walls of the solid to begin with.
Furthermore, the model geometry features dead-end pores at the top and bottom which are connected to the remaining structure by a single pore throat.
These cavities trap air bubbles that cannot be flushed out, and the respective void space becomes inaccessible to the fluids.
In the recorded images, these air bubbles appear similar to \ac{PDMS} as translucent material with a dark-gray to black contour, and they are manually removed.
Additionally, larger, non-persistent artifacts from the image that do not impact the actual experiment are accounted for by merging them with their surrounding~(\autoref{fig:mask_processing_tweaked}).
The images are further denoised by applying a median blur filter with a rectangle of kernel size $5 \times 5$, and thresholding subsequently converts it into a binary image.
Remaining small artifacts (below the minimum pore throat size) are further reduced by applying three iterations of the morphological operations \emph{opening} and \emph{closing} with a rectangular kernel of size $3 \times 3$.
Finally, contour extraction~\cite{suzuki1985topological} is applied by redrawing the binarized image and excluding all contours with an area $< 150$ pixels~(\autoref{fig:mask_processing_final}, cf.~\autoref{fig:segmentation_mask} for full image).
Parameters in the automatic part of the pre-processing were determined empirically; they were adjusted to reduce artifacts as effectively as possible without removing fine geometry or corrupting the topology of the pore network.

Input images are pre-processed automatically akin to the masks (without manual adjustment). 
With this, each image can be segmented into the three investigated phases~(\autoref{fig:segmentation_segmented}).
The solid phase corresponds to all pixels that are \textit{on} in the pre-processed, binarized mask.
The wetting phase consists of all those pixels which are \textit{on} in the binarized input images but \textit{off} in the respective mask.
All remaining pixels are accordingly assigned to the non-wetting phase.

\section{Visualization Data For Comparative Analysis}
\label{sec:comparability}

We now extract visualization data from the segmentation results that allow us to investigate processes quantitatively~(\autoref{sec:derived_values}).
Direct comparison between experiments further requires dealing with different temporal scales~(\autoref{sec:tempnorm}).

\subsection{Quantities of Interest}
\label{sec:derived_values}

\begin{figure}
  \includegraphics[width=\columnwidth]{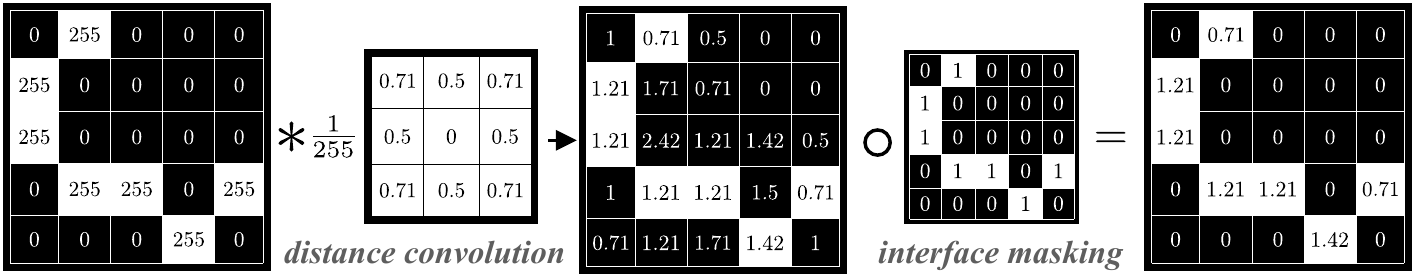}
  \caption{Determining the interface length.}
  \label{fig:interface_length}
\end{figure}

\paragraph*{Saturation of Fluid Phases}

Strictly speaking, the input images are two-dimensional projections of three-dimensional processes.
However, the loss in depth information of processes is negligible: the considered porous medium is orders of magnitude smaller in depth than in width and height, at \SI{100}{\micro\metre} in comparison to $\SI{20}{\milli\metre} \times \SI{15}{\milli\metre}$.
Accordingly, we can simply approximate the saturation of the fluid phases by the relative fraction of pixels per phase in relation to non-solid pixels in total.

\paragraph*{Interfacial Areas}
\label{sec:interfaces}
Given that depth can be neglected in our case, the interface between two phases is proportional to the summed length of the interface lines.
Our interface extraction generates one binary mask per phase combination, which contains one-pixel-thin lines at the position of the respective interface.
For this, the mask of the first phase is subtracted from a dilated version of itself (computed via a $3 \times 3$ rectangular kernel).
Then, a bitwise AND is applied to the result and the second considered phase.
The result then undergoes a series of filter operations that remove all those pixels that have both horizontal/vertical and diagonal neighbours (essentially removing staircase shapes along the interface line).

The interfacial area is then computed via the Euclidean distance between the centers of adjacent pixels~(\autoref{fig:interface_length}).
First, we calculate for each pixel the sum of distances from its center to the border of adjacent pixels on the interface via 2D convolution~(\textit{distance convolution}).
The entries of the employed $3 \times 3$ kernel accordingly exhibit the distances to the central element.
Second, we apply the respective interface mask, and determine the total interface length via the sum of pixel values.

\paragraph*{Disconnected Fluid and Inlet Connection}

A commonly employed assumption in the domain is that fluids do not get disconnected~(cf.\ \autoref{sec:related}).
However, we consider the explicit investigation of connected components to be of crucial importance for the analysis for a number of reasons.
The disconnection of the non-wetting phase during primary drainage is an effect that results from the synergy of the capillary number, the viscosity ratio, and the geometry of the porous medium.
For low capillary numbers, there is no specific reason for the disconnection of the non-wetting phase.
However, it is expected for high capillary numbers that the non-wetting phase will get disconnected due to the developed shear stresses, and their competition against interfacial forces, as the only means to absorb the energy fed into the system.
However, the location of these disconnections also depends on the local geometry of the porous medium, and how they can locally enhance the shear stresses due to local constrictions.
We expect that considering number and location of the disconnected non-wetting parts helps in quantifying the effect of the pore geometry on the evolution of flow for various boundary and physical conditions.
In addition to this, the extended theories of two-phase flow can be better evaluated, at later times, since the discrimination of the type of the newly formed interfaces with respect to the phases involved and their connectivity to the corresponding phase is an intrinsic characteristic of them.
To this end, we split the fluid phases into parts that are connected or disconnected from their respective inlet (e.g., on the right for the non-wetting phase, \autoref{fig:segmentation_segmented_detailed} and \subref{fig:segmentation_interface_detailed}).
We check which parts of fluid are connected by determining their distance to the respective image border where their inlet is located (a distance of 40 pixels in the 2448 pixels wide frame yielded reliable detection of both fluid parts across all experiments).
Disconnected components in the non-wetting phase correspond to blobs and ganglia, whereas we do not consider small droplets below 50 pixels~(blobs occupy a single pore, while ganglia stretch across pores).
We do not distinguish between blobs and ganglia here, but this is planned for future work.

\subsection{Temporal Normalization}
\label{sec:tempnorm}

Our analysis particularly focuses on the comparison of the experiments.
However, the experimental parameters have a strong influence on the speed at which the displacement processes take place, and accordingly experiments are recorded at different frame rates.
Since the image recordings are also started and stopped manually, there are no directly given reference points.
To make processes at different temporal scales directly comparable, we derive three reference points in time based on the flow of the non-wetting phase:  entry~$t_i$ of a phase into the domain, breakthrough~$t_b$ when an uninterrupted connection between the left and right boundary is formed~(\autoref{fig:expbreak}), and finally end $t_e$ when the experiment is stopped (this is done manually when all relevant processes have been captured).
Especially the universal and dimensionless time stamp for breakthrough is critical for further analysis: depending on the boundary conditions, an event, like the continuation of saturation change, may or may not happen after breakthrough.
In combination with the actual physical time, this provides the basis for categorizing experimental data.
Based on these reference points, we employ different temporal normalization schemes:
\emph{relative time}~$\hat{t}= \frac{t-t_i}{t_e - t_i} (\hat{t} \in [0,1])$, \emph{breakthrough normalization}~$\hat{t}= \frac{t-t_i}{t_b - t_i} (\hat{t} \in [0,1]$), and \emph{breakthrough scaling}~$\hat{t}= \frac{t-t_i}{t_b - t_i}$ ($\hat{t} \in [0,\frac{t_e-t_i}{t_b - t_i}]$, with fixed point $t_b=1$).
\emph{Relative time} has the advantage that it gives an impression of the speed with which the displacement processes take place.
The behaviour of the fluid before and after breakthrough is of particular interest in the analysis.
\emph{Breakthrough normalization} and \emph{breakthrough scaling} enables direct comparability of the experiments in this respect.

\def\chartWidth{0.3305\linewidth}
\begin{figure*}[t]
  \subcaptionbox{\label{fig:plot_w-n}Fluid Interface}{%
    \includegraphics[width=\chartWidth]{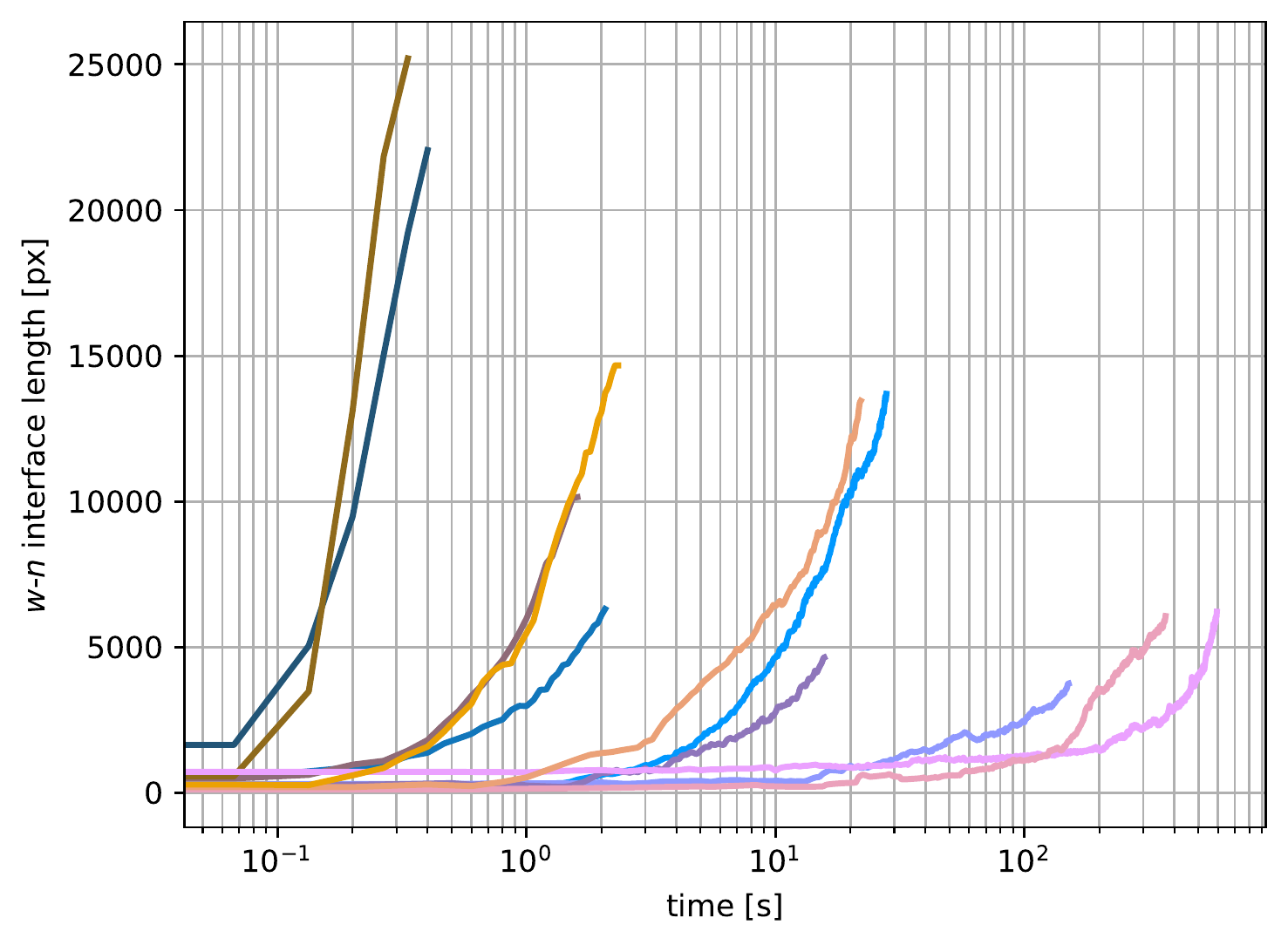}}
  \hfill
  \subcaptionbox{\label{fig:plot_S-cw}Connected wetting phase saturation}{%
    \begin{overpic}[width=\chartWidth]{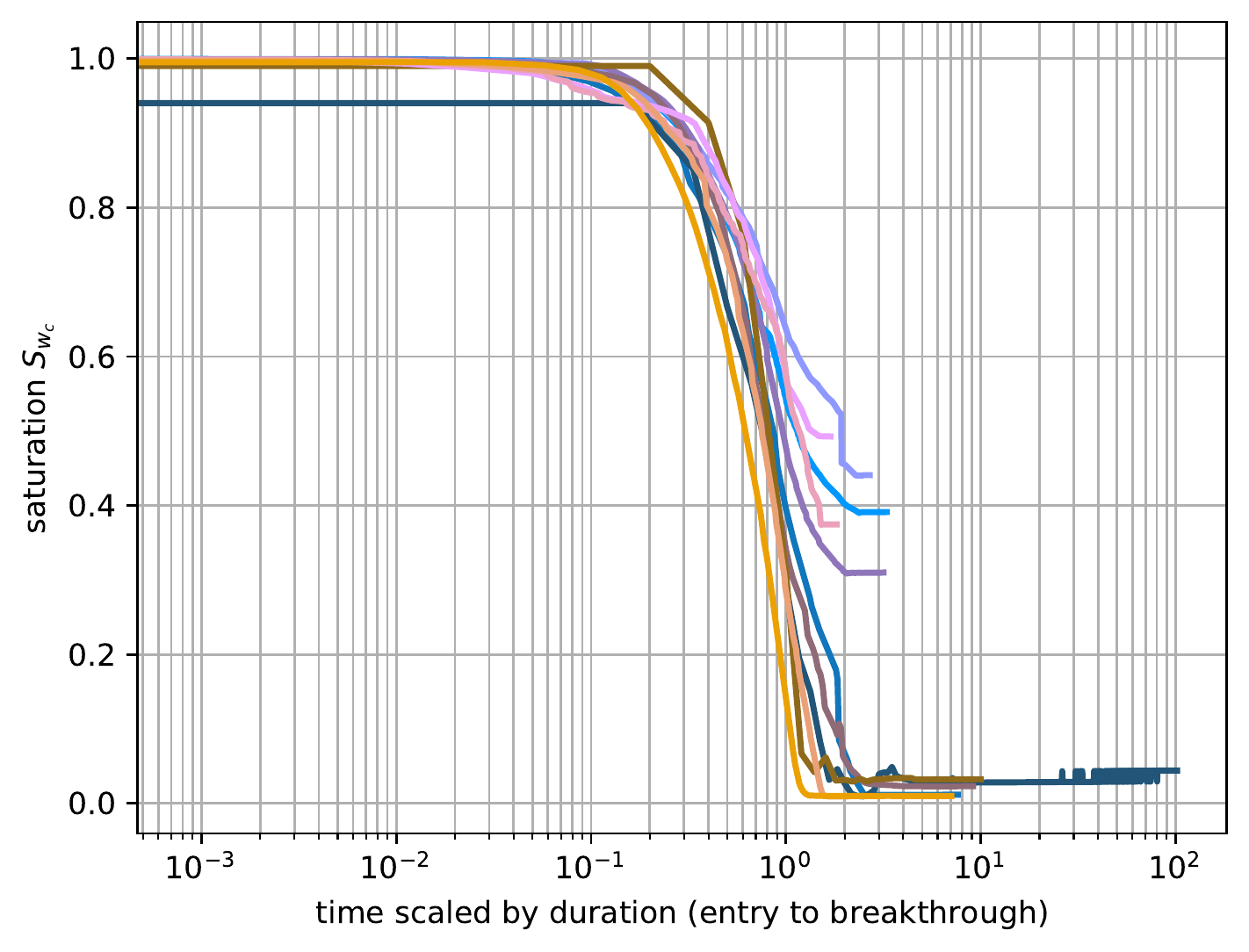}
      \put(12, 10){\includegraphics[scale=2.5]{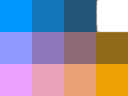}}
      \put(10.5, 39){\tiny\color{white}\parbox{.7cm}{\raggedleft $\ca\,\num{1e5}$ $\m\,0.2$}}
      \put(10.5, 27){\tiny\color{white}\parbox{.7cm}{\raggedleft $\ca\,\num{1e5}$ $\m\,1$}}
      \put(10.5, 15.5){\tiny\color{white}\parbox{.7cm}{\raggedleft $\ca\,\num{1e5}$ $\m\,10$}}

      \put(22.5, 39){\tiny\color{white}\parbox{.7cm}{\raggedleft $\ca\,\num{1e4}$ $\m\,0.2$}}
      \put(22.5, 27){\tiny\color{white}\parbox{.7cm}{\raggedleft $\ca\,\num{1e4}$ $\m\,1$}}
      \put(22.5, 15.5){\tiny\color{white}\parbox{.7cm}{\raggedleft $\ca\,\num{1e4}$ $\m\,10$}}

      \put(34.5, 39){\tiny\color{white}\parbox{.7cm}{\raggedleft $\ca\,\num{1e3}$ $\m\,0.2$}}
      \put(34.5, 27){\tiny\color{white}\parbox{.7cm}{\raggedleft $\ca\,\num{1e3}$ $\m\,1$}}
      \put(34.5, 15.5){\tiny\color{white}\parbox{.7cm}{\raggedleft $\ca\,\num{1e3}$ $\m\,10$}}

      \put(46.5, 27){\tiny\color{white}\parbox{.7cm}{\raggedleft $\ca\,\num{1e2}$ $\m\,1$}}
      \put(46.5, 15.5){\tiny\color{white}\parbox{.7cm}{\raggedleft $\ca\,\num{1e2}$ $\m\,10$}}

  \end{overpic}}
  \hfill
  \subcaptionbox{\label{fig:plot_n-blobsganglia}Number of blobs and ganglia}{%
    \includegraphics[width=\chartWidth]{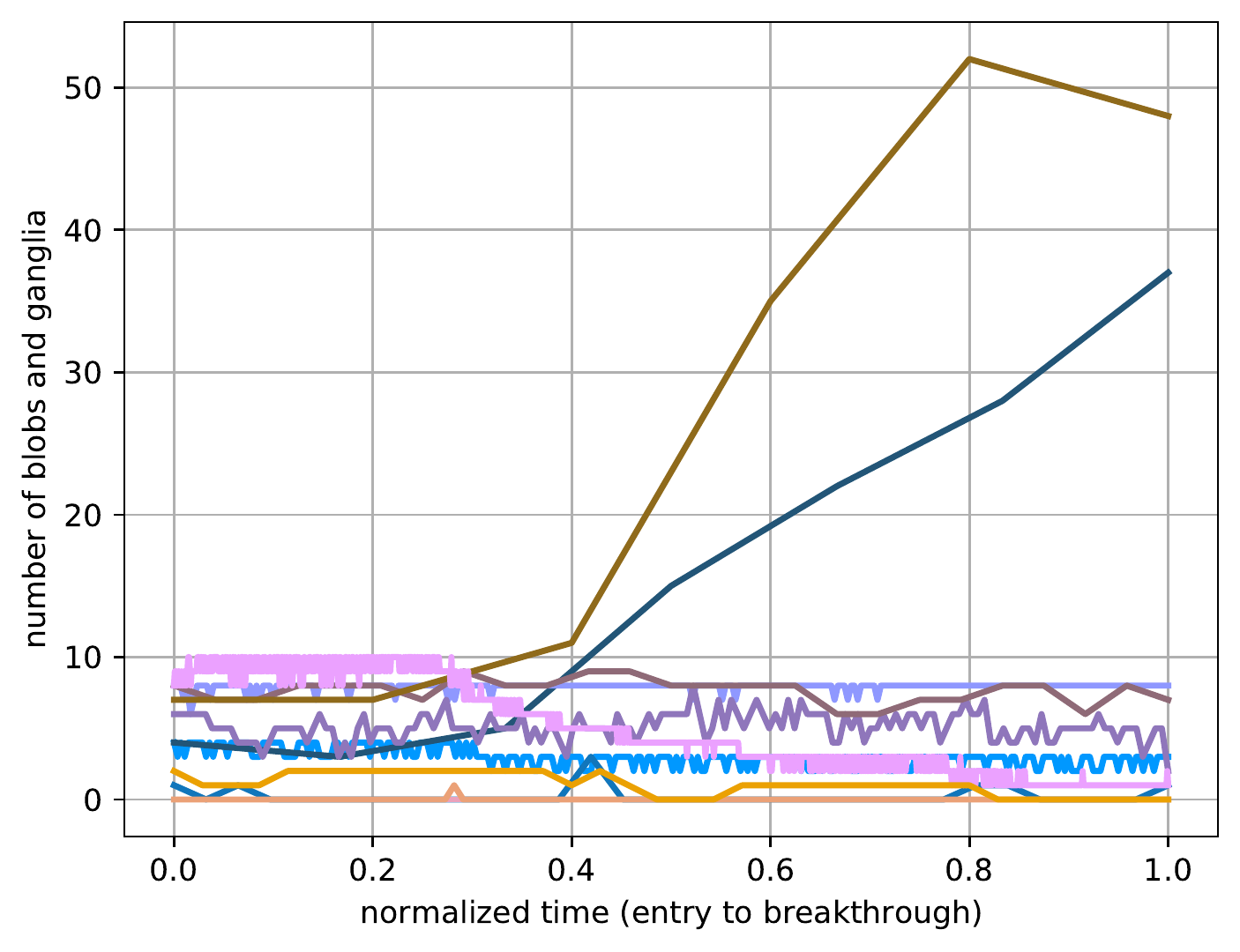}}
  \caption{Temporal evolution of different quantities of interest and varying modes for temporal normalization.}
  \vspace{-0.4cm}
  \label{fig:plot}
\end{figure*}

\section{Temporal Evolution of Characteristic Values}
\label{sec:charactersitic_values}

The extracted visualization data and different means to deal with temporal heterogeneity now allow us to quantitatively compare occurring processes across different experiment runs~(\autoref{fig:plot}).

\noindent
\textbf{Interface between wetting and non-wetting phase (\autoref{fig:plot_w-n})}
\noindent

Four groups of two to three curves can be identified, whereas the interface length starts increasing at similar points in time: 
\begin{compactitem}
\item $\{\textcolor{ca3m02}{\camns{3}{02}}, \textcolor{ca2m1}{\camns{2}{1}}\}$, 
\item $\{\textcolor{ca4m02}{\camns{4}{02}}, \textcolor{ca3m1}{\camns{3}{1}}, \textcolor{ca2m10}{\camns{2}{10}}\}$,
\item \{$\textcolor{ca5m02}{\camns{5}{02}}, \textcolor{ca4m1}{\camns{4}{1}}, \textcolor{ca3m10}{\camns{3}{10}}\}$,
\item and \{$\textcolor{ca5m1}{\camns{5}{1}}, \textcolor{ca5m10}{\camns{5}{10}}, \textcolor{ca4m10}{\camns{4}{10}}\}$.
\end{compactitem}
Interestingly, this means that a configuration of $\ca$ and $\m$ yields similar behavior in this regard to a configuration with $\ca$ being one order lower and $\m$ one order higher for the considered parameter range. 
It is also indicated that when we are in a capillary regime already (with $\textcolor{ca5m1}{\camns{5}{1}}$ and $\textcolor{ca4m10}{\camns{4}{10}}$), the next "step" of $\textcolor{ca5m10}{\camns{5}{10}}$ will still be capillary dominated, which means that the phase topology does not change and the resulting impact on the interface length is comparably small.
The investigation of respective transport networks that exhibit similar flow paths in \autoref{fig:grid_before_breakthrough} complements this observation.
Generally, it can be seen that higher capillary numbers yield a larger interfacial area, which can largely be attributed to increasing phase disconnection~(cf.~\autoref{fig:plot_n-blobsganglia}).
This effect is also one of the main reasons why we need to differentiate between connected and disconnected phases, since the extended theories and their validation only account for connected phases.

\noindent
\textbf{Saturation of the Connected Wetting Phase (\autoref{fig:plot_S-cw})}

\noindent
Two groups of curves can clearly be identified in the diagram: one where the saturation settles between 0.3 and 0.5 (\{$\textcolor{ca5m1}{(\num{1e-5},{1})}, \textcolor{ca5m10}{(\num{1e-5},{10})},
\textcolor{ca5m02}{(\num{1e-5},{0.2})},
\textcolor{ca4m1}{(\num{1e-4},{1}}),
\textcolor{ca4m10}{(\num{1e-4},{10}})\}$), 
and one where the saturation after the breakthrough approaches zero~(the rest).
This graph clearly reflects  that for increasing capillary number, the breakthrough saturation also increases.
In addition to this, for increasing viscosity ratios, the saturation change after breakthrough evolves smoothly and nearly undisturbed, resulting in a nearly total recovery of the wetting phase from the pore space~(the saturation of the connected wetting phase approaches zero).
These findings indicate the following behavior.
For low capillary numbers, where the flow regime is capillarity-dominated and capillary fingers make their appearance, as soon as breakthrough takes place, there is limited possibility for the saturation to change, since the connectivity of the wetting phase is also limited due to the two-dimensional nature of the porous medium, and topological changes are not favored.
As the capillary number increases, we gradually switch from a capillary regime to a mixed capillary and viscous regime.
The contribution of the viscous forces enables the re-mobilization of the the disconnected wetting phase, which eventually leads to an increased final saturation than this of breakthrough.
When we increase the capillary number even further (for viscosities $\m=0.2$ and $\m=1$), then blobs of disconnected non-wetting phase---caused by the competition between shear stresses and capillary forces---are rushed into the network (e.g.,~\textcolor{ca5m02}{\camns{5}{02}} in \autoref{fig:expimgs:ca5m02}, also see below in~\autoref{fig:plot_n-blobsganglia}).
This further increases the saturation, but in a erratic way due to the synergy with viscous fingering.
As the viscosity of the non-wetting phase also increases ($\m=10$), then we switch to a piston-like movement of the non-wetting phase, which invades a large part of the pore space with fingers~(i.e.,~elongated tail structures, e.g., \textcolor{ca2m10}{\camns{2}{10}} in \autoref{fig:expimgs:ca2m10}), establishing a steady state of increased saturation after breakthrough.

\noindent
\textbf{Disconnected Components (\autoref{fig:plot_n-blobsganglia})}

\noindent
We now consider the number of disconnected components (i.e., blobs and ganglia) between entry of the non-wetting phase and breakthrough.
This gives us a complementary view of the behaviour described my means of \autoref{fig:plot_w-n}.
In this diagram, the experiments  $\textcolor{ca2m1}{\camns{2}{1}}$ and $\textcolor{ca3m02}{\camn{-3}{0.2}}$ stand out, whose number is more than $\times 3$ above the values of the other experiments.
Accordingly, it shows that the combination of high capillary number and low viscosity ratio creates many more isolated and disconnected blobs of the non-wetting phase than the other configurations.
From these results, we would expect that for the configuration $\camn{-2}{0.2}$ we would see that the number of blobs would be even higher, since the high capillary number and the low viscosity ratio would favour such a behaviour.
This would have to to be confirmed with additional experimental measurements.
It is also clearly reflected that as the viscosity ratio increases for high capillary numbers, the system switches from creating blobs to creating thin fingers.
This leads to a corresponding increase of the interfacial area (as seen in \autoref{fig:plot_w-n}), but to a lesser degree since there is also a different mechanism creating this interfacial area.

\def\networkHeight{0.31\linewidth}
\begin{figure*}[t]
  \subcaptionbox{\label{fig:pore_networks} Illustration of transport network generation.}{%
    \adjincludegraphics[height = \networkHeight, trim={{0.\width} {0.0\height} {0.0\width} {0.05\height}}, clip]{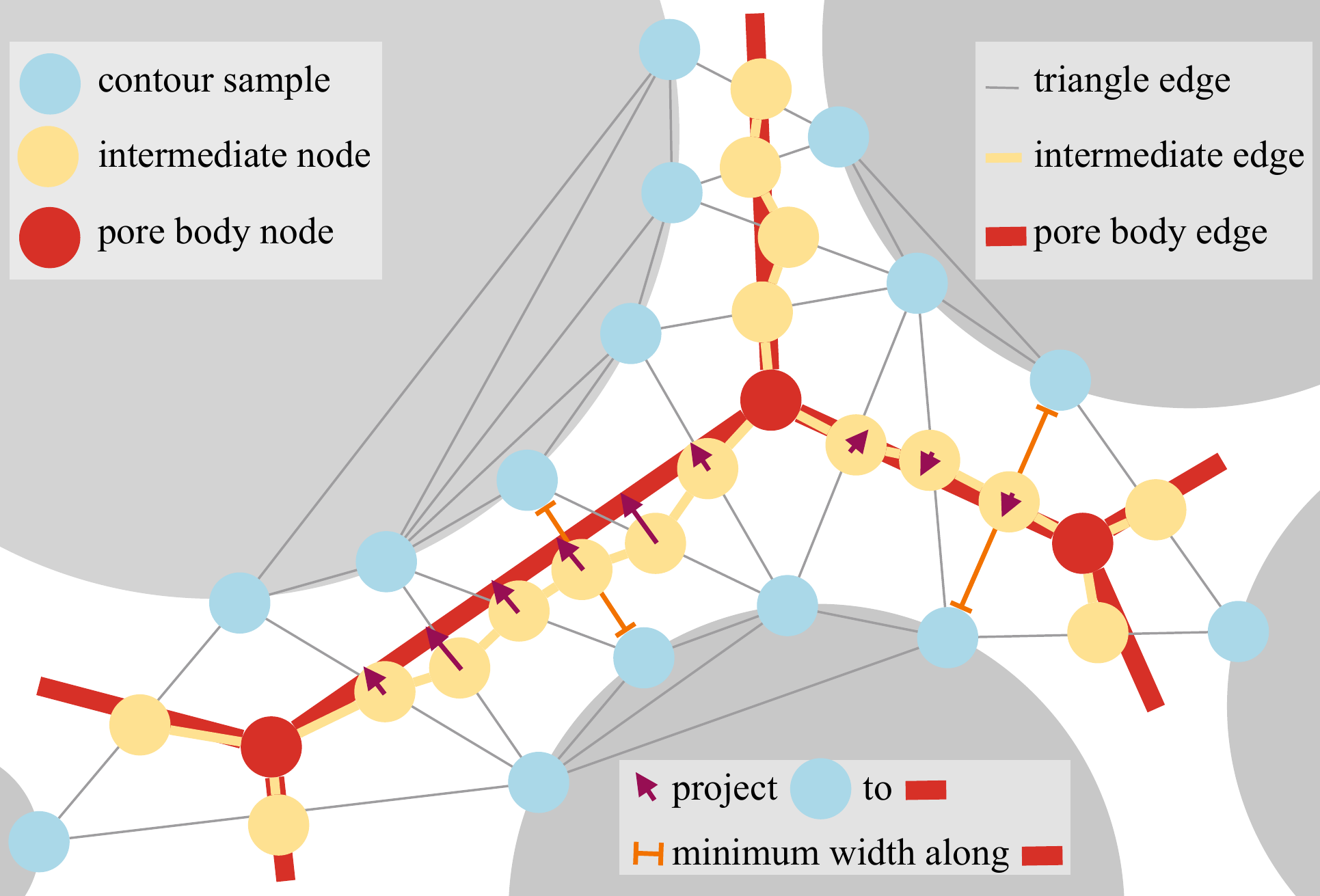}}
  \hfill
  \subcaptionbox{\label{fig:compute_network}result used for \autoref{fig:grid_before_breakthrough} and \autoref{fig:grid_after_breakthrough}}{%
    \adjincludegraphics[height = \networkHeight, trim={{0.\width} {0.1\height} {0.0\width} {0.1\height}}, clip]{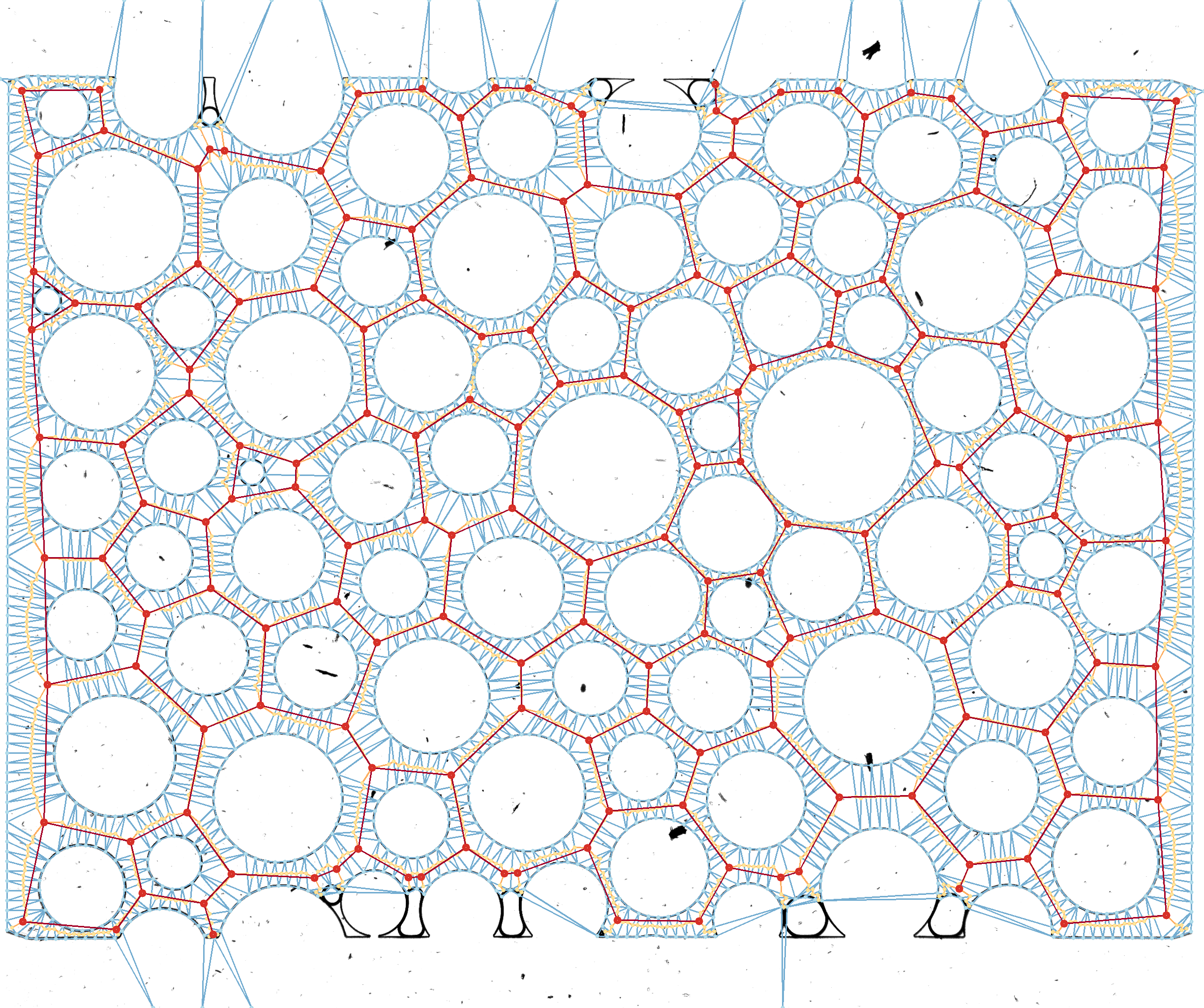}}

\caption{Generation of transport networks. (\subref{fig:pore_networks})~First, the contours of the solid are sampled (blue). Then, we generate a triangulation with these contour samples. Midpoints of edges not covered by solid become intermediate nodes (yellow). In the center of triangles with three intermediate nodes we insert a pore body node (red), while triangles with two intermediate nodes induce intermediate edges (yellow lines). These intermediate edges are then aggregated to form connections between pore body nodes (red lines). (\subref{fig:compute_network})~Result used in this paper.}
\vspace{-0.4cm}
\end{figure*}

\section{Space-Time Analysis via Transport Networks}
\label{sec:networks}

The analysis in \autoref{sec:charactersitic_values} is based on values of interest to reflect the state of the experiment at a given time.
While these values quantitatively show how $\ca{}$ and $\m{}$ affect the displacement processes, they do not convey how this impacts the resulting flow patterns and how they differ spatially between experiments.
For this, we introduce a compact representation based on node-link diagrams of so-called transport networks, which aggregate the data over time while maintaining spatial flow information.
The porous medium is modeled as a network of pore bodies (nodes) connected by pore throats (links).
This concept is widely used in the estimation of material properties and the simulation of different processes in porous media research; however, no precise definition of pore bodies and throats exists~\cite{blunt2001flow, xiong2016review}.
We developed an approach based on the triangulation of the whole domain to not only extract topological information but also capture covered pore and throat areas.

\subsection{Transport Network Visualization}
\label{sec:network_generation}
We first create a graph capturing the pore network structure w.r.t. the solid~\textbf{(1)}.
In doing this, we associate each node and edge (i.e., pore and pore throat) with a certain region in the original data.
This then allows to determine the fluid coverage for each edge and node across experiments and time steps~\textbf{(2)}.

\textbf{(1)\,Extraction of porous structure network graph.}
Transport networks and their topology are based on a triangulation of the void space, i.e., the part of the porous medium that is available for fluid transport (cf.~\autoref{fig:pore_networks}).
For this, we consider the extracted mask interfaces~(cf.~\autoref{sec:interfaces}).
Each contour is described by a clockwise ordered list of coordinates of the associated pixels which are sampled by selecting indices with a fixed step size~(blue points in \autoref{fig:pore_networks}).
In our application, a step size of 20 pixels has proven to be sufficient to adequately resolve the domain.
From this set of points, an unstructured triangle grid is generated by computing a Delaunay triangulation.

For each triangle edge, we then consider its midpoint and check whether it lies in void space.
For this we use the preprocessed, binary mask image of the micro-model, as described in \autoref{sec:imageanalysis}. %
To compensate for rounding errors due to converting the calculated floating point positions of the midpoints into integer pixel coordinates, the midpoints are checked against a version of the mask image where the solid phase is dilated by one pixel.
If the midpoint lies in void space it becomes an intermediate node (yellow circle in~\autoref{fig:pore_networks}) and is added to our transport network.
We then consider the triangles again and check for the numbers of intermediate nodes at their edges.
In case there are two intermediate nodes, we add an intermediate edge connecting them to our transport network (yellow line in~\autoref{fig:pore_networks}).
If there are three intermediate nodes, a new pore body node is inserted at the barycenter of the triangle (red circle in~\autoref{fig:pore_networks}).
In addition, connections of the three intermediate nodes to the pore body node are inserted as well.
Thus, a pore throat connecting two pore bodies is formed by at least two intermediate edges.

\textbf{(2)\,Checking for fluid progression.}
Each pore body node and intermediate node is associated with one triangle each which again represents a portion of the porous medium that is available for fluid transport.
Each of these elements is indexed via $n$ below.
A high coverage of the area of a triangle by the transported fluid indicates that the corresponding element of the network is used for fluid transport.
The normalized coverage $\mathcal{C}_{n, t}=\tfrac{A_{n, t}}{A_{\Delta_n}}$ is computed as the ratio between triangle area $A_{n, t}$ and the area of the full triangle $A_{\Delta_n}$
Based on this, we now associate a usage rate $\mathcal{U}$ with each intermediate edge and pore body node of the network.
We define this measure $\mathcal{U}_{n}$ as the number of time steps $t$ at which the normalized coverage of the triangle of element $n$ is larger than a fixed threshold $k$, divided by the total number of time steps of the investigated time period $T$:
\begin{equation}
  \label{eq:usage_rate}
  \mathcal{U}_{n} = \frac{\abs{\{t | t \in T \land \mathcal{C}_{n, t} > k\}}}{\abs{T}}.
\end{equation}
In our analysis we set a comparably low threshold of $k = 0.1$ to exclude small dust particles in the micro-model from being detected as transported fluid.
However, the sensitivity w.r.t. $k$ is rather low, and our measurements showed only little impact for $k\in [0.1, 0.7]$.

\textbf{(3)\,Visual representation.}
For visualizing the transport network, intermediate line segments between two pore bodies are contracted to form a single edge connecting pore bodies (red line in \autoref{fig:pore_networks}).
Their thickness is scaled linearly with the minimum width of the represented pore throat.
We approximate this width by the minimum length of corresponding triangle edges~(orange in \autoref{fig:pore_networks}).
The \emph{usage rate} $\mathcal{U}$ of each segment is mapped onto the contracted edge by orthogonally projecting its intermediate nodes onto the single edge and assigning the value to this segment (projection indicated by purple arrows).
Finally, the contracted edge is drawn segment by segment, mapping the usage rate to color using the \emph{Inferno} color map\cite{njsmith2015}).
Complementing $\mathcal{U}$, an important characteristic of interest for the analysis is the length of the considered time interval.
To incorporate this information, we use unoccupied pores and throats with $\mathcal{U}<0.01$ (i.e, that would otherwise be mapped to black in our case).
For this, we use a secondary isoluminant color map based on \emph{CET-I1} by Kovesi\cite{kovesi2015good} (which we consider to be rather low-key visually compared to Inferno).

An aspect of particular interest is how the flow paths change after the breakthrough, but to clearly show that the transport network in \autoref{fig:grid_after_breakthrough} is created in a slightly modified way.
In the first step, the usage rate is calculated for the complete time interval after breakthrough until the end of the recording.
In the second step, these results are compared to the usage rate values before breakthrough and mapped to black for any element that was used for fluid transport before breakthrough ($\mathcal{U} > 1 \%$).

The generated void space triangle mesh, nodes and connections can be applied to all experiments that have been conducted with the same micro-model, after they have been aligned as described in \autoref{sec:comparability}.
This allows to analyze and compare the experiments with a uniform representation, clearly conveying differences in flow processes.

\subsection{Transport Network Matrix}
\label{sec:network_glyphs}
\begin{figure}[t]
  \centering
  \DeclareDocumentCommand{\naframe}{mm}{figs/20200429_networks/Ca=10#1,M=#2/transport-network/after_breakthrough}
  \def\nImgWidth{0.234\linewidth}
  \subcaptionbox*{\label{fig:grid_after_breakthrough:ca5m02}\color{ca5m02} $\ca\,\num{1e-5}, \m\,0.2$}%
                 {\includegraphics[width=\nImgWidth,cfbox=ca5m02 1pt 0pt]{\naframe{-5}{0.2}}}
                 \hfill
                 \subcaptionbox*{\label{fig:grid_after_breakthrough:ca4m02}\color{ca4m02} $\ca\,\num{1e-4}, \m\,0.2$}
                                {\includegraphics[width=\nImgWidth,cfbox=ca4m02 1pt 0pt]{\naframe{-4}{0.2}}}
                                \hfill
                                \subcaptionbox*{\label{fig:grid_after_breakthrough:ca3m02}\color{ca3m02} $\ca\,\num{1e-3}, \m\,0.2$}
                                               {\includegraphics[width=\nImgWidth,cfbox=ca3m02 1pt 0pt]{\naframe{-3}{0.2}}}
                                               \hfill
                                               \subcaptionbox*{}
                                                              {\includegraphics[width=\nImgWidth,cfbox=white 1pt 0pt]{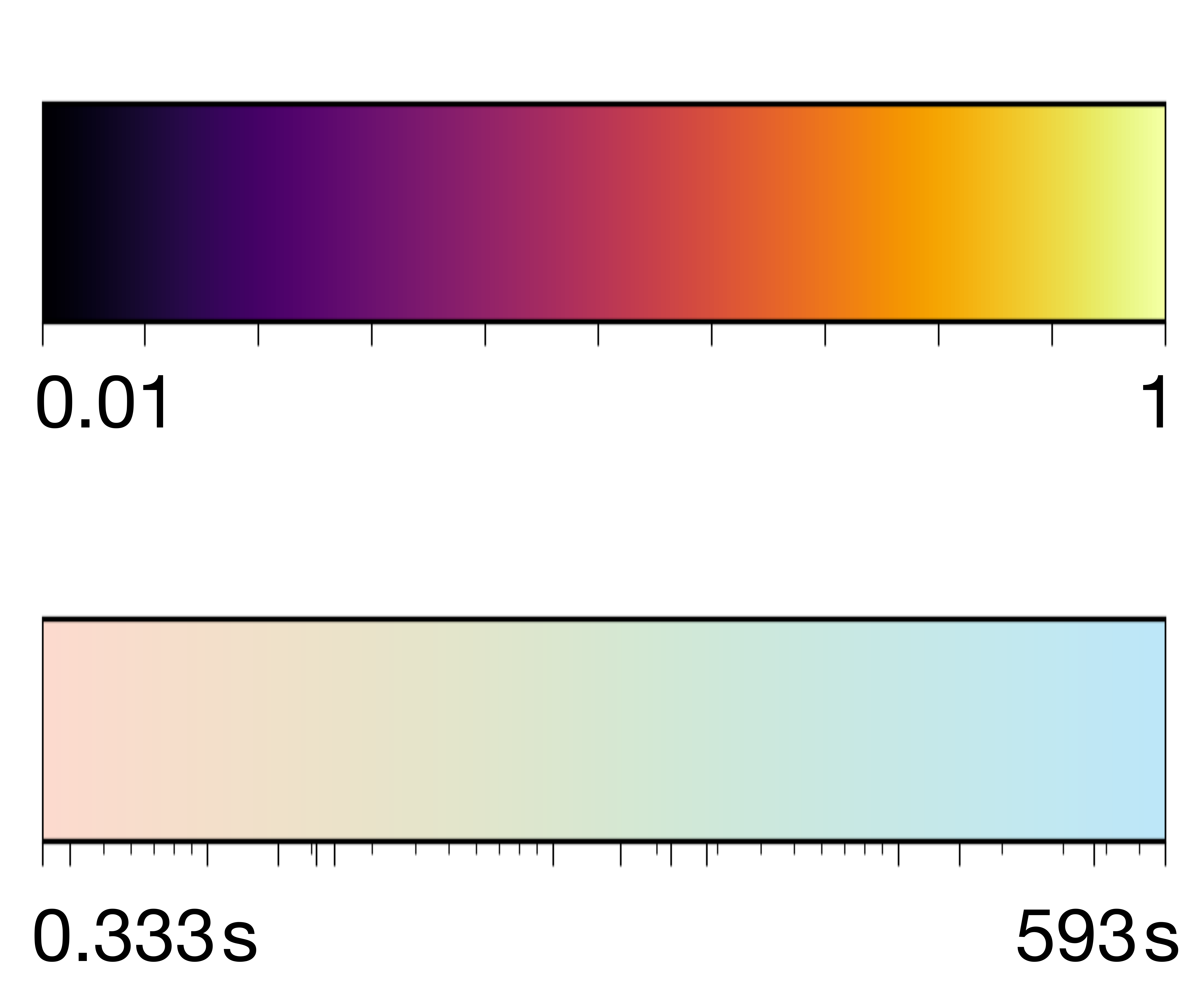}}

                                                              \subcaptionbox*{\label{fig:grid_after_breakthrough:ca5m1}\color{ca5m1} $\ca\,\num{1e-5}, \m\,1$}%
                                                                             {\includegraphics[width=\nImgWidth,cfbox=ca5m1 1pt 0pt]{\naframe{-5}{1}}}
                                                                             \hfill
                                                                             \subcaptionbox*{\label{fig:grid_after_breakthrough:ca4m1}\color{ca4m1} $\ca\,\num{1e-4}, \m\,1$}
                                                                                            {\includegraphics[width=\nImgWidth,cfbox=ca4m1 1pt 0pt]{\naframe{-4}{1}}}
                                                                                            \hfill
                                                                                            \subcaptionbox*{\label{fig:grid_after_breakthrough:ca3m1}\color{ca3m1} $\ca\,\num{1e-3}, \m\,1$}
                                                                                                           {\includegraphics[width=\nImgWidth,cfbox=ca3m1 1pt 0pt]{\naframe{-3}{1}}}
                                                                                                           \hfill
                                                                                                           \subcaptionbox*{\label{fig:grid_after_breakthrough:ca2m1}\color{ca2m1} $\ca\,\num{1e-2}, \m\,1$}
                                                                                                                          {\includegraphics[width=\nImgWidth,cfbox=ca2m1 1pt 0pt]{\naframe{-2}{1}}}

                                                                                                                          \subcaptionbox*{\label{fig:grid_after_breakthrough:ca5m10}\color{ca5m10} $\ca\,\num{1e-5}, \m\,10$}%
                                                                                                                                         {\includegraphics[width=\nImgWidth,cfbox=ca5m10 1pt 0pt]{\naframe{-5}{10}}}
                                                                                                                                         \hfill
                                                                                                                                         \subcaptionbox*{\label{fig:grid_after_breakthrough:ca4m10}\color{ca4m10} $\ca\,\num{1e-4}, \m\,10$}
                                                                                                                                                        {\includegraphics[width=\nImgWidth,cfbox=ca4m10 1pt 0pt]{\naframe{-4}{10}}}
                                                                                                                                                        \hfill
                                                                                                                                                        \subcaptionbox*{\label{fig:grid_after_breakthrough:ca3m10}\color{ca3m10} $\ca\,\num{1e-3}, \m\,10$}
                                                                                                                                                                       {\includegraphics[width=\nImgWidth,cfbox=ca3m10 1pt 0pt]{\naframe{-3}{10}}}
                                                                                                                                                                       \hfill
                                                                                                                                                                       \subcaptionbox*{\label{fig:grid_after_breakthrough:ca2m10}\color{ca2m10} $\ca\,\num{1e-2}, \m\,10$}
                                                                                                                                                                                      {\includegraphics[width=\nImgWidth,cfbox=ca2m10 1pt 0pt]{\naframe{-2}{10}}}
                                                                                                                                                                                      \caption{
                                                                                                                                                                                        Transport networks of non-wetting flow after breakthrough~(structures with fluid before breakthrough are depicted in black; color maps are akin to~\autoref{fig:grid_before_breakthrough}).
                                                                                                                                                                                      }
                                                                                                                                                                                      \label{fig:grid_after_breakthrough}
\end{figure}

For investigating the effects of $\ca{}$ and $\m{}$ on flow behaviour, we create a small multiples visualization that arranges transport networks such that it reflects their position in the parameter space.
Showing all available combinations of capillary numbers and viscosity ratios allows to directly compare their impact on the spatio-temporal distribution of phases in a static visualization.
In our analysis, we investigate two kinds of transport matrices, differing in the considered time span: before (\autoref{fig:grid_before_breakthrough}) and after (\autoref{fig:grid_after_breakthrough}) breakthrough.
Their combination conveys a comprehensive qualitative and quantitative summary of the evolution of flow, and also allows to identify isolated effects and evaluate their contribution to phase distribution.

First, we consider the time span from entry to breakthrough in \autoref{fig:grid_before_breakthrough}.
Generally, the progression of the flow through the porous medium can clearly be seen, especially also indicating for how long certain structures were occupied.
It is of particular interest to see that late in the experiment new flow paths are covered close to the inlet of the model (as indicated via low usage rate~$\mathcal{U}$), with a residence time similar to the occupied areas close to the outlet.
This means that these effects occur later than could be expected, which happens to some extent in all experiments, but is particularly significant for larger $\ca$ and $\m$.
This provides an indication of the importance of the phase occupancy at the entry pores of the flow network in these cases especially, and how they contribute to the breakthrough saturation.
We further clearly see that for low capillary numbers~($\can{-5}$),  the viscosity ratio~$\m$ does only have little impact on the distribution of phases before breakthrough, as they show approximately the same spatial distribution.

From~\autoref{fig:grid_after_breakthrough}, it becomes apparent in which experiments the flow paths still change significantly after breakthrough.
This behavior indicates a flow dominated by viscous rather than capillary forces, and allows us to partition the parameter space into different regimes.
For low capillary numbers $\ca$, we see that after breakthrough the usage rate does not change significantly since the capillary forces dominate the process, and the geometry of the porous medium is in charge of the process.
The minor changes which take place can mostly be attributed to so-called capillary end effects, which would not be of essence in a bigger porous medium (capillary end effects arise from the discontinuity of capillarity in the wetting phase at the outlet).
For intermediate capillary numbers, we observe some similarities with the previous case, in the sense that the distribution of phases changes so as to increase the breakthrough saturation~(but not drastically).
However, there is a stronger tendency toward newly visited pores close to the inlet.
This is the outcome of the effect of the more dynamic boundary conditions in the network of channels outside of the field of view, which were used to feed the porous medium with the non-wetting phase.
We also discover significant mobilization of the non-wetting phase at later times which increased the final saturation.
We attribute this effect to the fact that the non-wetting phase has reached the outlet of the model and re-mobilization was inevitable due to extensive viscous effects.
As the capillary number is increased even further to $\can{-2}$, we see the breakthrough saturation to increase accordingly, with extensive re-mobilization of the non-wetting phase due to the increased viscous effects.
Crucially, the re-mobilization of the non-wetting phase covers (nearly) the whole pore space, which is an important property for numerous applications.
Another interesting aspect we learn from the visualization is the effect of the phase occupancy at the boundary pores.
The degree of occupancy mostly depends on the boundary conditions, but the visualization shows that for high capillary numbers---i.e. in a viscous flow regime---the breakthrough saturation and the phase distribution are both highly affected by the fact that the occupancy of boundary pores varies significantly.
This can impact the comparability of the experiments.
Resolving this issue by adapting the experimental setup is almost impossible in practice since for high capillary numbers even a small deviation in terms of surface roughness or pump fluctuations can have a significant effect.
This emphasizes the importance of showing such deviations as a crucial part of a comprehensive analysis.

\begin{figure}[t]
  \centering
  \begin{overpic}[width=\columnwidth]{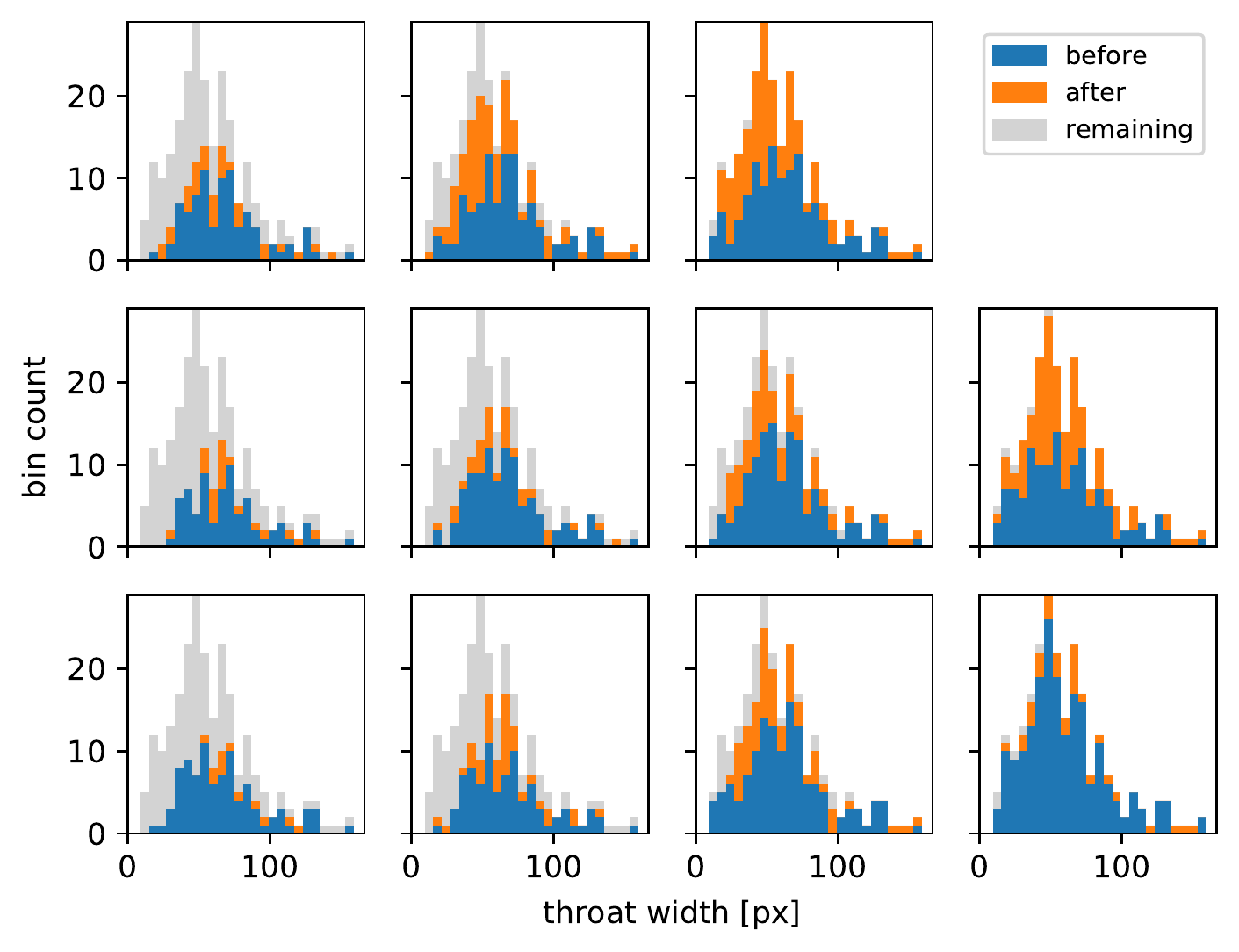}
    \put(19, 70){\color{ca5m02}\parbox{.7cm}{\raggedleft $\ca\,\num{1e5}$ $\m\,0.2$}}
    \put(19, 47){\color{ca5m1}\parbox{.7cm}{\raggedleft $\ca\,\num{1e5}$ $\m\,1$}}
    \put(19, 24){\color{ca5m10}\parbox{.7cm}{\raggedleft $\ca\,\num{1e5}$ $\m\,10$}}

    \put(42, 70){\color{ca4m02}\parbox{.7cm}{\raggedleft $\ca\,\num{1e4}$ $\m\,0.2$}}
    \put(42, 47){\color{ca4m1}\parbox{.7cm}{\raggedleft $\ca\,\num{1e4}$ $\m\,1$}}
    \put(42, 24){\color{ca4m10}\parbox{.7cm}{\raggedleft $\ca\,\num{1e4}$ $\m\,10$}}

    \put(65, 70){\color{ca3m02}\parbox{.7cm}{\raggedleft $\ca\,\num{1e3}$ $\m\,0.2$}}
    \put(65, 47){\color{ca3m1}\parbox{.7cm}{\raggedleft $\ca\,\num{1e3}$ $\m\,1$}}
    \put(65, 24){\color{ca3m10}\parbox{.7cm}{\raggedleft $\ca\,\num{1e3}$ $\m\,10$}}

    \put(88., 47){\color{ca2m1}\parbox{.7cm}{\raggedleft $\ca\,\num{1e2}$ $\m\,1$}}
    \put(88., 24){\color{ca2m10}\parbox{.7cm}{\raggedleft $\ca\,\num{1e2}$ $\m\,10$}}
  \end{overpic}
  \caption{
    Histograms over minimum width of pore throats that transported the non-wetting phase before and after breakthrough; cf.~\autoref{fig:grid_before_breakthrough} (before) and \autoref{fig:grid_after_breakthrough} (after) for corresponding networks.}
  \label{fig:grid_histogram}
\end{figure}

\subsection{Pore Throat Histogram}
\label{sec:histograms}
Abstracting the porous medium as a network allows us to characterize the displacement process by discrete paths through which the non-wetting phase flows.
It further crucially allows to create histograms that make a connection between pore throat width and fluid transport.
In the presentation of the histograms, we differentiate between pore throats that already transported fluid before the breakthrough and those that only do so after the breakthrough (akin to the analysis in~\autoref{sec:network_glyphs}).
The two types are visualized as stacked bars with blue and orange color, respectively.
The remaining pore throats, which do not transport fluid, are added on top as light-gray bars to convey the total distribution of pore throat widths in the porous medium.

These histograms in \autoref{fig:grid_histogram} complement the transport network visualization in \autoref{fig:grid_before_breakthrough} and \autoref{fig:grid_after_breakthrough}, covering both time periods (before/after breakthrough) in one view and allowing for a more quantitative comparison of throat occupation.
Different aspects already discovered in the transport network visualizations can be investigated in more detail.
Generally, larger $\ca$ lead to a higher saturation, with $\ca\in\{\num{1e-3}, \num{1e-2}\}$ yielding (almost) full saturation of the porous medium eventually. This complies with the general understanding that high capillary numbers in combination with low viscosity ratios lead to viscous fingering, low capillary numbers with practically any viscosity ratio lead to capillary fingering, and high capillary numbers in combination with high viscosity ratios lead to a stable front. 
Here, it can be seen that the viscosity parameter $\m$ crucially impacts when parts of the medium are saturated: with lower $\m$ large portions of the domain are only saturated after breakthrough.
It is also confirmed that---across all experiments---a lower coverage is achieved with smaller throat widths, due to the increased entry capillary pressure of the corresponding throats in a drainage process.

\section{Discussion and Future Work}
\label{sec:conclusion}

\label{sec:analysis}
\label{sec:discussion}

In this work, we conduct an in-depth visual analysis of the characteristics of two-phase  fluid flow and its preferred transport paths based on new experimental data.
In particular, we study the characteristic flow patterns resulting from different combinations of viscous and capillary forces, and the impact of the porous medium structure.
For this, we extract quantities of interest from the experiments with the high accuracy required, and employ breakthrough as a reference point to achieve comparability between processes running at different time scales across experiments.
This enables quantitative analysis, but abstracts spatial effects.
To additionally consider this aspect, we introduce a new spatio-temporal visual representation to depict which paths through the porous medium were occupied to what extent.
Apart from visualizing it directly, it can also be useful to extract abstract information regarding spatial aspects, namely the occupancy of throats depending on their width.

Identifying and classifying experiments with similar behavior on this basis allows us to partition the experiment parameter space into regimes and relate it to the dominant driving forces.
An important insight gained from the visual analysis is that we can now see that the reason for the creation of the interfacial area is completely different in two characteristic cases.
One is due to the breakup of the non-wetting phase due to the competition between the viscous and capillary forces, while the other one is due to the creation of thin fingers.
Based on our conducted experiments, we can now not only quantify the two different effects, but we can also spatially correlate them, and map them to the pore space using our spatio-temporal representation.
There are significant implications driven by this ability to differentiate between degenerative identical states in a robust, compact and efficient way, since these findings can be attributed to different effects, which correspond to different approaches regarding the physical context.
In general, our experiments exhibit physical effects that are challenging to model due to the increased physical complexity of the processes involved.
While micro-scale approaches exist, they cannot adequately model all observed effects, whereas macro-scale models do not consider some effects at all (e.g., the reduction of effective interfacial area between solid and non-wetting fluid not appropriately reflected). 
Based on the extracted data and findings of this work, we plan to make a first attempt to effectively model the respective processes.
We aim to investigate the role of the solid phase geometry more closely, especially in the interplay with the boundary and physical conditions of the fluids involved.
We also intend to extend our research to scenarios further than primary drainage, and evaluate the role of the pore bodies and throats of the porous structure as the dominant geometrical measure of each displacement process.
This is directly linked to the Young-Laplace equation, where capillary pressure is dominantly affected by the local geometrical properties of the porous medium.
Pore bodies are the determining feature for drainage, due to their increased size and lower entry capillary pressure for an invasion process, while pore throats are dominant during imbibition due to their reduced size and the spontaneous response of the system to the increased capillarity.
Additionally, the role of the disconnected phase will be investigated more closely, both in terms of their effective influence in the evolution of flow as a pressure barrier, but also in terms of the local, boundary, and physical conditions under which a phase can become disconnected.

These efforts will require a much more densely sampled experiment parameter space (via further experiments or conducted simulations), posing additional challenges for the analysis as well.
In particular, our visualization approaches should also be made more scalable visually, and we aim to consider different directions in this regard.
First of all, while our transport networks proved to be crucial in our analysis, they cannot directly scale visually to larger experiment and/or simulation ensembles.
However, they already provide a simplified representation of the pore structure, that could further be scaled towards more expressive smaller representations (like glyphs) by reducing the underlying graph accordingly.
In addition, we aim to conduct further extraction of other quantities that could be relevant (e.g., distinguishing between ganglia and blobs and quantifying their shape via descriptors).
This accordingly results in a high-dimensional feature vector describing each experiment, that could further be the basis for further analysis techniques based on projection and/or clustering.

Furthermore, in this work, we employ the same porous media model for the sake of direct comparability.
A larger, more comprehensive study, however, could include different models for a more targeted investigation of specific effects.
This would be supported by our full pipeline: the porous model for the experiments was generated using AutoCAD (and could easily be adapted) and all other steps, including the transport network visualization, could be done in exactly the same way.
However, naturally, direct spatial comparison across different pore structures is not possible directly anymore, but needs to be substituted by more abstract alternatives.
Perspectively, 3D experiments (and simulations) could also be conducted for further investigation.
The extraction of quantities generally works akin to the 2D case, and the transport networks can be generalized to 3D as well: based on the discretization of the porous medium by a tetrahedral mesh, triangle edge midpoints are simply replaced with the barycenters of the triangular tetrahedron faces, and midpoint checks can also be adapted accordingly in a straight-forward fashion.

\section*{Acknowledgements}
We thank the German Research Foundation~(DFG) for funding this work within SFB 1313, Project Number 327154368.

\bibliographystyle{eg-alpha-doi}
\bibliography{template,nikos,gleb,porousFlow}

\end{document}